\documentclass[a4paper,fleqn,usenatbib]{mnras}

\usepackage{graphicx}	% Including Figure files
\usepackage{amsmath}	% Advanced maths commands
\usepackage{amssymb}	% Extra maths symbols
\usepackage{multicol}        % Multi-column entries in tables
\usepackage{bm}		% Bold maths symbols, including upright Greek
\usepackage{pdflscape}	% Landscape pages
\usepackage{graphicx}
\usepackage{lipsum}
\usepackage{tikz}
\usetikzlibrary{matrix,shapes,snakes, arrows,positioning,chains}
\usetikzlibrary{shapes,fit}

\newcommand{\ks}{\mbox{$K_{\rm s}$}}
\usepackage{calrsfs}

\newcommand{\arcs}{\hbox{$^{\prime\prime}$}~}
\newcommand{\degree}{\mbox{$^{\circ}$}}

\usepackage{amsmath,bm,times}

\usepackage[T1]{fontenc}
\usepackage{ae,aecompl}

\usepackage{newtxtext,newtxmath}
\title[Optimal correction of distortion for High Angular Resolution images]{Optimal correction of distortion for High Angular Resolution images. Application to GeMS data}

\author[A. Bernard]{A. Bernard$^{1}$\thanks{Contact e-mail: \href{mailto:anais.bernard@lam.fr}{anais.bernard@lam.fr}}\thanks{Present address: 38, rue Fr\'ed\'eric Joliot-Curie 13388 Marseille cedex 13 FRANCE}, B. Neichel$^{1}$, L. M. Mugnier$^{2}$ , T. Fusco$^{2}$,$^{1}$
\\
$^{1}$Aix Marseille Universit\'e, CNRS, LAM (Laboratoire d'Astrophysique de Marseille) UMR 7326, 13388 Marseille, France \\
$^{2}$ONERA \- The French Aerospace Laboratory, F-92322 Chatillon, France}
\pubyear{2017}

\begin{document}
\label{firstpage}
\pagerange{\pageref{firstpage}--\pageref{lastpage}}
\maketitle

% Abstract of the paper
\begin{abstract}

Whether is ground-based or space-based, any optical instrument suffers from some amount of optical geometric distortion. Recently, the diffraction-limited image quality afforded by space-based telescopes and by Adaptive Optics (AO) corrected instruments on ground based-telescope, have increased the relative importance of the error terms induced by optical distortions. In particular, variable distortions present in Multi-Conjugated Adaptive Optics (MCAO) data are limiting the astrometry and photometry accuracy of such high resolution instruments. 
From there, the ability to deal with those phenomenon had become a critical issue for high-precision studies. 
We present in this paper an optimal method of distortion correction for high angular resolution images. Based on a prior-knowledge of the static distortion the method aims to correct the dynamical distortions specifically for each observation set and each frame.
The method follows an inverse problem approach based on the work done by \cite{Gratadour2005} on image re-centering, and we aim to generalized it to any kind of distortion mode. The complete formalism of a Weighted Least Square minimization, as well as a detailed characterization of the error budget are presented. In particular we study the influence of different parameters such as the number of frames and the density of the field (sparse or crowed images), of the noise level, and of the aliasing effect. Finally, we show the first application of the method on real observations collected with the Gemini MCAO instrument, GeMS/GSAOI. The performance as well as the gain brought by this method are presented.

\end{abstract}

% Select between one and six entries from the list of approved keywords.
% Don't make up new ones.
%\begin{keywords}
%adaptive optics -- high angular resolution -- image processing 
%\end{keywords}
%
%%%%%%%%%%%%%%%%%%%%%%%%%%%%%%%%%%%%%%%%%%%%%%%%%%

%%%%%%%%%%%%%%%%% BODY OF PAPER %%%%%%%%%%%%%%%%%%

% The MNRAS class isn't designed to include a table of contents, but for this document one is useful.
% I therefore have to do some kludging to make it work without masses of blank space.
\begingroup
\let\clearpage\relax
\tableofcontents
\endgroup
\newpage

\section{Introduction}

All optical instruments, whether is ground-based or space-based, suffer from some amount of optical geometric distortion.
These distortions have multiple origins that induce different behavior.
They may be static, resulting from unavoidable errors in the optics shape and placement  or from imperfect fabrication of the detectors. For exemple, in the Wide Field Planetary Camera 2 (WFPC2) of the Hubble Space Telescope (HST), each of its four CCDs suffers from a distortion that reaches 5 pixels at the edge of the field (\cite{Anderson2003}). 
Distortions may be dynamic, due for exemple, to environmental parameters such as temperature variation: a typical illustration is the short term focus changes caused by small motions of the Hubble Space Telescope (HST) Optical Telescope Assembly (OTA) secondary mirror (referred to as OTA \textit{breathing} in \citet{Hasan1994}).
In Adaptive Optics (AO) corrected images from ground based telescopes, dynamical distortions may also be coming from residuals from the AO correction, or, in the case the AO system and the imaging camera are mounted at the Cassegrain focus of the telescope, from gravity flexure of the instrument as the telescope tracks (\cite{Cameron2009}).

The diffraction-limited image quality afforded by space-based telescopes and by AO corrected instruments on ground based-telescopes, increases the relative importance of the error terms induced by these phenomena. A large number of studies evidence that the largest instrumental systematic that limits the astrometry accuracy in any optical system is the geometric distortion \citep{Anderson2003, Cameron2009, Trippe2010, Fritz2010}. As the goal of astrometry is to measure the position of target stars over many epochs, it is usually done on combined objects frames with varying pointing  positions (technique of \textit{dithering}), and varying epochs. The knowledge of the distortion is then necessary to place stellar positions in a globally correct reference frame. 
Distortion can also seriously limit photometric accuracy on stack images that are essential to photometer the faintest objects (\cite{Bernard2016}). 
Indeed, if it is stated that the angular
resolution on stacked images is always worse that the angular resolution on a single
frame, stacking images uncorrected for distortion amplified this degradation proportionally to the amount of present distortion.
%Indeed, stacking images uncorrected for distortion degrades the resolution proportionally to the amount of present distortion. 
For exemple, in the case of images collected with the Gemini Multi-Conjugated Adaptive Optics System (GeMS), the final angular resolution on a stack of images uncorrected for distortion can reach almost twice the typical angular resolution on a single frame (which is about 90\,mas).

The importance of distortion effects in high angular resolution imaging has led researchers to develop a number of strategies to mitigate their impact. 
%If geometric distortions are stable, then a number of strategies can be employed to mitigate their effect. 
%One method is to model the distortion to high accuracy;
The most notable method is the self calibration developed by \cite{Anderson2003} to derive with high accuracy (to $\sim$\,1\,milliarcsecond (mas)) the distortion of the HST WFPC2. The method consists in using the instrument to calibrate itself. The standard way to do this is to take multiple observations of the same field at different offsets and orientations, so that the same patch of sky is imaged on different regions of the detector. Observed positions for all the stars in all the images are measured, and a single distortion model is found. 
This method dispenses the need for an external reference frame, as the distortion-free reference positions are derived from the observations themselves. However, this requires a large set of on-sky data with large translations at many orientations to constrain all high-order modes of distortion. For exemple, the solution derived by  \cite{Anderson2003} required 80 exposures and 4000 stars per images. This kind of calibration, which cannot be implemented for each observation, suppose that the distortion is stable in time. It is well adapted to space-based telescopes where the weightless environment means that telescope flexure does not lead to large changes in the distortion solution. Thus, this method has next been used for numerous HST cameras: the Advanced Camera for Survey High Resolution Channel (ACS/HRC) with a precision of 0.25\,mas (\cite{Anderson2004}), the UVIS channel of the WFPC3 with a precision of 1\,mas (see \cite{Bellini2009}) and the ACS Wide Field Channel (WFC) with a precision of 1\,mas (\cite{Ubeda2013}). Later, \cite{Anderson2006} adapted this solution for ground-based telescopes by calibrating the Wide Imager Field (WIF) at the European Southern Observatory 2.2\,meter telescope with a precision of 7\,mas (on each coordinate). Following the same process, a distortion solution was derived for the High Acuity Wide field K-band Imager (HAWK-I) at the Nasmyth focus of Unit Telescope 4 / Very Large Telescope (UT4/VLT) ESO 8 meter telescope with a precision of 3\,mas (\cite{Libralato2014}), and for the Large Binocular Cameras-blue, at the prime focus of the Large Binocular Telescope with a precision of 15\,mas (\cite{Bellini2010}).

Another way to solve for the distortion present in an instrument is to observe a field where we have prior knowledge of the reference positions of all the stars in a distortion-free system. Distortion would then show itself immediately as the residuals between the observed and the reference positions of stars. 
Reference positions can be derived from external astrometric reference frames, such as astrometric catalogs (UCAC-2, GSC-2 or 2MASS) or HST observations (accurately corrected from distortion by self-calibration as mentioned previously) as it was done by \cite{Yelda2010} and \cite{Service2016} to calibrate the Near-InfraRed Camera 2 (NIRC2) on the Keck II 10\,meters telescope. The cross-calibration might also be done using datasets from other wavelength regimes. A nice example is given by \cite{Reid2007} who used precise Very-Long-Baseline Interferometry (VLBI) astrometry of SiO maser stars in order to define an astrometric reference frame in K-band images of the Galactic center. The major advantage of this method is that a much smaller set of on-sky data is needed as compared to the self-calibration method. The final error on the distortion calibration is then set by the residual distortion in the distortion-free external reference used.  
 \\
 The difficulty to calibrate distortion on ground-based telescopes is due to the variability of the distortions. Some are quite stable, as the NIRC2 on the Keck II telescope which system was stable over the period from 2007-2010 (\cite{Yelda2010}), but have to be calibrated again after modifications in the system such as realignment (\cite{Service2016}), and some are varying in time. Thus, the solution derived for the WIF/ESO, varied of about 100\,mas in 3\,years and the one derived for HAWK-I/VLT varied of about 3\,mas on 3\,months scale. This phenomenon is particularly presents in Multi-Conjugate Adaptive Optics (MCAO) systems where the distortion may vary during one night (\cite{Neichel2014a}, \cite{Massari2016}). The most likely hypothesis on the origins of these variable distortions is that they might be introduced by the deformable mirrors conjugated in altitude.
 In this context, several authors have performed a relative correction of the distortion based on a master-coordinate-frame created by averaging the position of each star over all frames and considered as a reference. With this method, the correction can adapt to each set of data and each single frame, but the reference is not absolute. The measurement of the relative astrometric precision achieved with this method is then calculated as the Root-Mean-Square (RMS) deviation of the positions through the different frames. 
 %It ranges between 1.2 and 2.8\,mas for the MCAO Demonstrator MAD (\cite{Meyer2011}) while \cite{Neichel2014a} achieve with this method, a relative astrometry precision of 0.4 mas on GeMS data. 
The relative astrometric precision  ranges from 1.2 to 2.8\,mas for MCAO Demonstrator MAD
(\cite{Meyer2011}) and reaches 0.4\,mas on GeMS/GSAOI data \cite{Neichel2014a}.

Finally, some \textit{in-lab} methods are developed to characterize the instrument distortion with dedicated measurements in the laboratory or at the telescope. One example is the "north-south test" used for the spectro-imager SINFONI at the VLT (e.g. \cite{Abuter2006}). This method uses devices that illuminate the detectors with well-defined images or light patterns. Comparing the theoretical with the observed images allows for a description of the distortion.
Another exemple is the use of a calibration mask located in the focal plane of the imager. Such a mask could be a regular pattern of holes in an opaque material. 
The precision of the holes positions in the calibration mask determines the accuracy of the distortion calibration. For exemple, distortions in NIRC2 were initially characterized using illuminated pinhole masks (\cite{Cameron2007a}). However, the residual distortion in those solutions was larger than the distortion solution produced using on-sky method (\cite{Service2016}). 

In this paper, we present a new method of distortion correction
based on a prior-knowledge of the static distortion
as a starting point to correct the residual distortion (referred hereafter as dynamical
distortion). Thus, the correction to apply is calculated specifically for each frame of each observation set. Based on a \textit{Weighted Least Squares} (WLS) minimization, this method has the particularity to provide an estimation of both a distortion-free reference and a distortion solution associated to each frame with a limited noise propagation thanks to the \textit{weighted} property of the minimization. Both estimated parameters can be used independently depending on the scientific aim of the study: for exemple, absolute astrometry requires a good estimation of the distortion-free reference, whereas a good estimation of the distortion solution associated to each frame is needed for relative astrometry applications (that require an good ability to place stellar positions in a common reference frame) and for photometry studies (that require a good ability to stack the images). The performance of these estimations depends on the nature of the data (crowded or sparse fields, number of frames, noise level) and are detailed in Section \ref{sec:simu}.
The paper is constructed as follow : In Section \ref{sec:model}, we expose hypothesis on the data construction and on the noise statistical model. Section \ref{sec:algo} is dedicated to the description of the minimization process, and of the algorithm implementation. In Section \ref{sec:simu}, we describe the validation of the proposed method using simulated data, and Section \ref{sec:gems} shows its first application on on-sky data. 
%The data are collected with the Gemini MCAO instrument, GeMS combined with the Infra-Red (IR) camera GSAOI (for Gemini South Adaptive Optics Imager). 
The performance as well as the gain brought by this method are presented in this section.
%First, the off-axis parabola present in the AO bench as well as the deformable mirror conjugated in altitude are introducing low spatial orders of distortions. Then, from the analysis carried in Neichel et al. (2014b) \cite{Neichel2014b}, it has been showed that higher spatial orders are also present (around 15 degrees of freedom). Part of these high order distortions might be due to the infrared camera and errors in the gaps correction. Indeed, the camera is composed of four distinct detectors separated by gaps. During the data reduction process, the four detectors are considered to be perfectly positioned, which might not be the case in reality. During the dithering of the telescope, a star moving from one detector to another between the different frames might introduce high orders of distortion. Finally, another source of errors comes from dynamical distortions, which depends on the Natural Guide Stars (NGS) constellation and environmental factors like the telescope pointing and the dithering. 
%The static distortions can be calibrated by using distortion-free reference frames, as HST data Therefore, the correction of distortion must take into account a large number of degree of freedom and be specific to each epoch. 
\section{Data construction}
\label{sec:model}
%We investigate a new distortion correction tools that will minimize the distortion error on stacked high resolution images. The minimization of this error term will then increase the astrometric and photometric performance during the scientific analysis. 
The distortion correction method presented in this paper follows an \textit{inverse problem} approach based on the work developed by D. Gratadour  and L. Mugnier  in \cite{Gratadour2005}. Their original work was focusing on isoplanatic image recentering while we generalize it to any kind of distortion in the field. 
The inverse problem approach requires the modeling of the data formation process (which is called the \textit{direct problem}), in order to take it into account during the inversion. 
The direct problem provides the calculation of the data associated to an known object, it is thus used to build simulated data. This section is dedicated to the description of this direct problem, including the modeling of the distortions and the description of the statistical noise properties considered.

\subsection{Distortion modeling}

The object of interest considered here is the distortion-free reference frame $I^{ref}$ that would be delivered by a distortion-free instrument.
The associated data is then the image, noted $I^{data}$, delivered by a real instrument. The operation that connects the object to the data is a distortion function noted $D$, that affects coordinates:
%$\mathcal{D}$:
\begin{equation}
I^{data}[\,.\,,\,.\,]=I^{ref}[D(\,.\,,\,.\,)]
\end{equation}
This expression adapts to each frame $i$, with its associated distortion function. 
%It can apply directly to pixels, or to remarkable points in the fields. 
%In particular, it can apply to the center of gravity of remarkable objects present in the image, as galaxies or stars. In the following, we use the position of stars as remarkable points to describe the method: the position of a star $j$ in the frame $i$ is noted $X_{i,\,j}^{data}$. 
It can apply to any basis: from the most exhaustive one, the pixels themselves, to a subset of it such as a set of as local positions computed on specific point like source objects. Hereafter, we consider a basis of stars positions, where the position of star $j$ in frame $i$ is noted $X_{i,\,j}^{data}$.
Following the previous reasoning, any star position measured in the data, $X_{i,\,j}^{data}$, can be described as the position of the star $j$ in the reference frame $X_{j}^{ref}$, on which the distortion function $D_i$ is applied. 
\begin{equation}
X_{i,\,j}^{data}=D_i\left(X_{j}^{ref}\right)
\end{equation}
From here, different functions can be used to describe the distortion. 
Most commonly, 2D polynomials up to about 3rd or 5th order, depending on the authors, are used as distortion models. For exemple, \cite{Anderson2003} used a 3rd order polynomial fitting to correct distortion on the HST/WFPC2, while \cite{Libralato2014} used a 5th order polynomial fitting for the HAWK-I calibration. 
The NIRC2 camera at the W.M. Keck Observatory, was successively corrected for distortion using a polynomial fitting (see \cite{Ghez2008}, and \cite{Lu2009}), a bivariate B-spline fitting (\cite{Yelda2010}) and more recently a 2D Legendre polynomials fitting (\cite{Service2016}). The last authors compare the efficiency of different models: 2D cartesian polynomials, bivariate B-spline and 2D Legendre polynomial. They conclude that the 2D Legendre polynomial basis provides a faster convergence and lower residuals. 
%With this approach, no prior knowledge on the geometry of the problem is needed, however, a large number of model parameters have to be calculated. If a physical model for the distortion is at hand, a corresponding model can be substantially simpler (e.g. \cite{Trippe2008} ).
%A recent example for this approach is the analysis of the nuclear star cluster of the Milky Way by \cite{Trippe2008} and \cite{Gillessen2009}). They use the 3rd-order model to correct the distortion of the imager NAOS/CONICA at the VLT. In cases where analytic solutions are not feasible or not accurate enough, empiric descriptions might be used instead or in addition. This means that the information of interest is stored in look-up tables. This approach is necessary if significant high-frequency distortion is present. In a detailed analysis of the WFC/HST, \cite{Anderson2002} uses a combined “polynomial model plus lookup-table” to model the distortion of the camera. 
Following this study, we choose the 2D Legendre polynomial basis, which combines several advantages, to define our distortion model. First, the 2D Legendre polynomial basis is an orthogonal basis (e.g. \citet{dunkl2014orthogonal}, \citet{Ye2014} ) defined on a square, thus well adapted to describe the distortion on square images. Then, the ascending polynomial degree organization of the basis is convenient to characterize the distortion using a limited number of modes. 
Each Legendre polynomial $P_n(x)$ is an $n$th-degree polynomial and may be expressed in one dimension as follows:
% According to previous study (\cite{Service2016}) we choose the Legendre polynomiasl basis defined in one dimension, as follows :
\begin{equation}
P_n(x)=\frac{1}{2^n}\sum\limits_{k=0}^{n}\begin{pmatrix}n\\k\end{pmatrix}^2(x-1)^{n-k}(x+1)^{k}, ~x \in [-1;1]
\end{equation}
\noindent In addition, we choose to normalize the polynomials so that each mode contains the same amount of distortion. The final 2D distortion basis $B$ is defined as follows :
\begin{equation}
B_{N_\text{modes}}(x,y)= \begin{pmatrix}L_0(x).L_0(y)\\ L_1(x).L_0(y)\\L_0(x).L_1(y) \\L_2(x).L_0(y)\\L_1(x).L_1(y)\\L_0(x).L_2(y)\\...\end{pmatrix}=\begin{pmatrix}\frac{1}{2}\\ \frac{\sqrt{3}}{2}x\\\frac{\sqrt{3}}{2}y \\ \frac{\sqrt{5}}{4}(x^2-1) \\\frac{3}{2}xy\\\frac{\sqrt{5}}{4}(y^2-1)\\...\end{pmatrix} ~(x,y) \in [-1;1]^2\end{equation}
where:\\
$L_n$ is the $n$th degree normalized Legendre polynomial:
$L_n=\frac{P_n}{\mid \mid P_n\mid \mid}$
with
$\mid \mid P_n \mid \mid^2=\frac{2}{2n+1}$

%produit scalaire
%left\langle P_n,P_n\right\rangle

%\end{equation}
\noindent$ N_\text{modes} $ is the number of modes considered. It determines the dimension of the vector $B_{N_\text{modes}}$.\\

\noindent Let us recall that the 2D Legendre polynomial basis conserves the orthogonality property with respect to the inner product $\left\langle\,.\,,\,.\,\right\rangle$ defined on  
%$\mathbb{R}$ 
$[-1,1]^2$ as follows:
\begin{equation}
\left\langle f,g\right\rangle=\int_{-1}^{1}\int_{-1}^{1}f(x,y).g(x,y)\,dxdy\\
\end{equation}

~\\
\noindent Hereafter, the distortion modes are referred to using their index mode ($m$) defined as: 
 \begin{equation} 
m(k,l)=l(l+1)/2+k~~~~~~~~~~~~~~\text{and}\\
%\end{equation}
% \noindent and
 %\begin{equation} 
b_{m}(x,y)=L_{l-k}(x)*L_{k}(y)
\end{equation}
\noindent where $0 \leq k \leq l \in \mathbb{N}$ and $0 \leq m \leq N_\text{modes}$.\\

\begin{figure}
\includegraphics[width=\hsize]{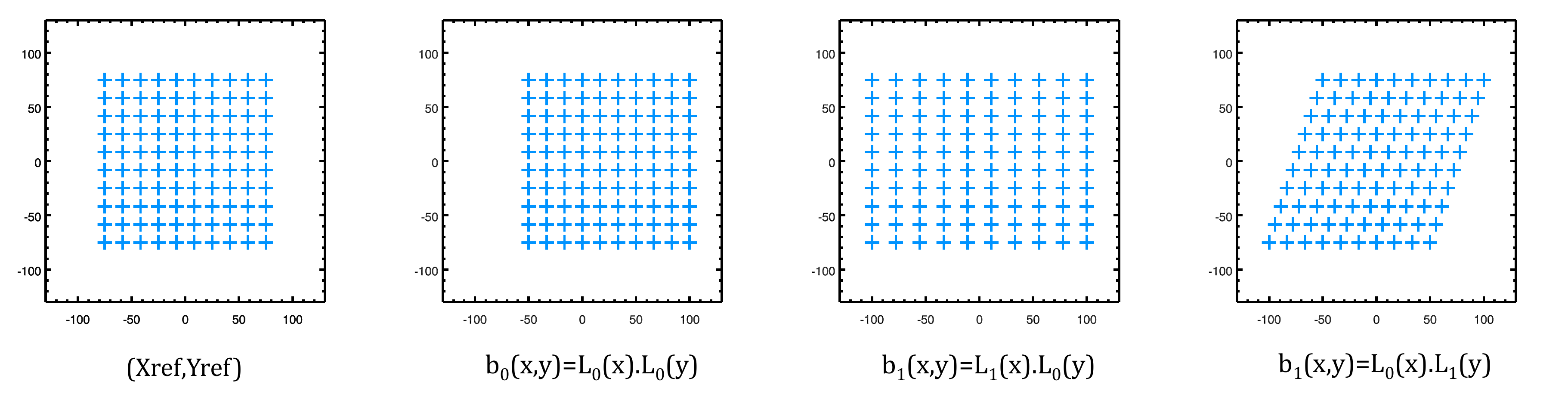}
\caption{Displacements induced by the three first modes of the distortion basis $B$, described by the 2D normalized Legendre polynomials. The displacements are applied to the $x$-coordinates. The left panel shows the reference grid without any distortion.}
\label{Figure:leg}
\end{figure}

\noindent Figure \ref{Figure:leg} shows the displacements induced by the three first modes of the distortion basis $B$, described by the 2D normed Legendre polynomials. The displacements are applied to the $x$-coordinates. The left panel shows the reference grid without any distortion. 

\noindent Finally, the position of a star $j$, on which a distortion $D_i$ is applied can be written as follows: 
%\begin{equation}
%D_i[X_{j}^{ref}]=\begin{pmatrix}x^{ref}_j\\y^{ref}_j\end{pmatrix}+\begin{pmatrix}a_{0,x} & a_{1,x} & ... & a_{N_\text{modes}-1,x}  \\a_{0,y} & a_{1,y} & ... & a_{N_\text{modes}-1,y}\end{pmatrix} B_{N_\text{modes}}\big(x^{ref}_j,y^{ref}_j\big)
%\end{equation}t

\begin{equation}
%\label{eq:model}
\resizebox{\hsize}{!}{$
X_{i,\,j}=D_i[X_{j}^{ref}]=\left\{ \begin{array}{ll} x^{ref}_j+\sum\limits_{l=0}^{d}  \sum\limits_{k=0}^{l} a_{i,~m,~x}~b_{m}(x_j^{ref},y_j^{ref})\\ 
y^{ref}_j +\sum\limits_{l=0}^{d}  \sum\limits_{k=0}^{l} a_{i,\,m,\,y}\,b_{m}(x_j^{ref},y_j^{ref}) \end{array}\right.$}
\end{equation}
\noindent where:\\
%\noindent  $m(l,k)=l(l+1)/2+k$ and  $0 \leq m \leq N_\text{modes}$.\\ 
\noindent $d$ is the maximal polynomial order considered. Note that $N_\text{modes}=(d+1)(d+2)/2$. \\

\begin{figure}
\includegraphics[width=\hsize]{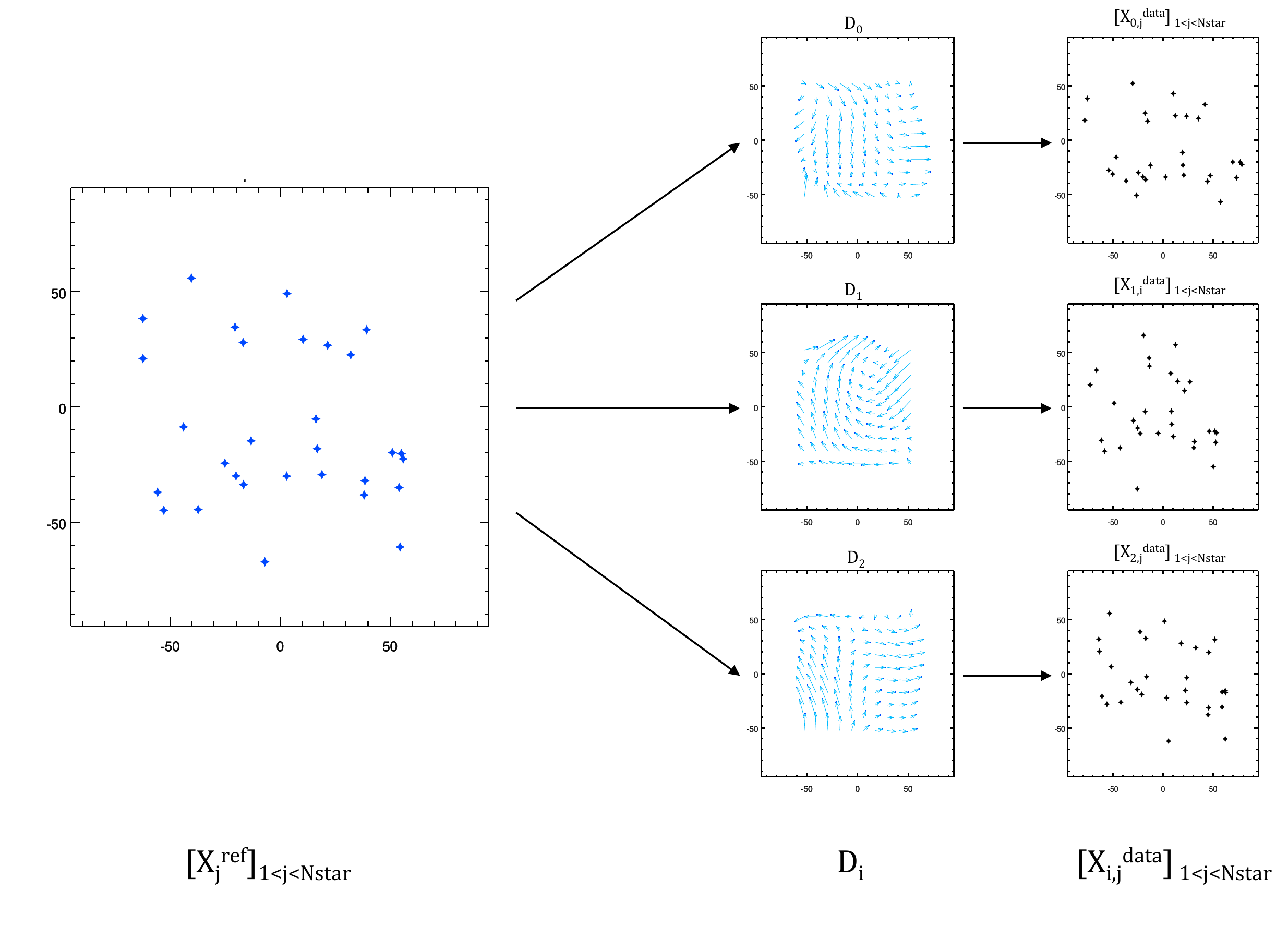}
\caption{Schematic diagram of the distortion application on different images. Left panel: reference postions $[X^{ref}]_{1 \leq j \leq N_{star}}$. Middle panel: set of distortion maps, $D_i$ (where $i=0,1,2$). Right panel: resulting positions after application of the associated distortion map $[{X^{data}}_{i,\,j}]_{1 \leq j\leq {N_{star}}}$.}
\label{Figure:distoschem}
\end{figure}

\noindent For each image, the set of decomposition coefficients,  $\big(a_{i,\,m,\,y/x}\big)_{0\leq m \leq N_\text{modes}}$ is noted as $A_i$ and referred to as the \textit{distortion coefficients}. The application of the distortion on different images is shown in Figure \ref{Figure:distoschem}. The left panel shows the reference positions $[X^{ref}]_{1 \leq j \leq N_{star}}$, with $N_{star}$ the number of reference sources, the middle panel shows the displacement induced by a set of distortion maps, $D_i$ (with $i=0,1,2$) and the right panel shows the distorted positions associated to each distortion map $[{X^{data}}_{i,\,j}]_{1 \leq j\leq {N_{star}}}$.

\subsection{Noise statistics}
\label{sec:noise}
The statistical properties of the noise that we should consider to
build the data model must be consistent with the noise present in measurement of star positions. 
%We assume here that this noise follows a white Gaussian law distribution. 
We assume here a noise that follows a Gaussian
white distribution (Gaussian White Noise).
This hypothesis is motivated by the following statements:
first, the images are contaminated by two main type of noise: the detector noise, which follow a Gaussian distribution, and the photon noise which follows a Poisson distribution. However, it has been shown thats for a high level of flux, a Poisson distribution converge towards a Gaussian distribution (see e.g. \citet{Mugnier2004}).
Then, the position's measurements combine informations of many pixels. Following the central limit theorem, the noise present in these measurements tends toward a normal distribution. As the noise present in each measurement is independent, we can consider with a very good approximation that the noise present in the data is white and Gaussian, with a variance noted $\sigma_{meas}^2$ that may depend on image number $i$ and star number $j$. The final data model, a.k.a. direct model, is thus: 

\begin{equation}
\label{eq:model}
\resizebox{\hsize}{!}{$
X_{i,\,j}=D_i[X_{j}^{ref}]=\left\{ \begin{array}{ll} x^{ref}_j+\sum\limits_{l=0}^{d}  \sum\limits_{k=0}^{l} a_{i,~m,~x}~{L_{l-k}(x_{j}^{ref})}{L_{k}(y_{j}^{ref})} + n_{i,\,j,\,x}\\ 
y^{ref}_j +\sum\limits_{l=0}^{d}  \sum\limits_{k=0}^{l} a_{i,\,m,\,y}\,{L_{l-k}(x_{j}^{ref})}{L_{k}(y_{j}^{ref})}+ n_{i,\,j,\,y} \end{array} \right.$}
\end{equation}
where $n_{i,\,j,\,x}$ and $n_{i,\,j,\,y}$ are independent Gaussian noises, with variance $\sigma_{meas,\,i,\,j}$, of the position measurements.

%%%%-------------------------  ALGO DESCRIPTION ----------------------------------------------------------------------------------

\section{Algorithm description}
\label{sec:algo}
In the previous section we defined the direct model. To solve the distortion problem, we now aim to inverse this model through the minimization of a criterion. In this section, we define the final criterion to be minimized and describe the implementation of the resulting algorithm. Then, we present the estimated parameters utilisations, and the error criteria associated. 

\subsection{Definition of a criterion}

The final direct model is described by Equation (\ref{eq:model}). In this equation, both the distortion coefficients $A_i$, and the reference positions $X^{ref}_j$, are unknown. The proposed method aims to perform a joint estimation of these two parameters for all the stars $j$ ( with $j \leq N_{star}$ and $N_{star}$ the number of stars considered) and all the frames $i$ (with $i \leq N_{im}$ and $N_{im}$ the number of frames). We introduce here the following notations:
\begin{itemize}
\item[$\cdot$]$\big[X^{ref}\big] =\big[X^{ref}_{j}\big]_{1\leq j\leq N_{star}}$, the whole set of reference positions;  
\item[$\cdot$]$\big[A~\big]= \big[A_{i}\big]_{1\leq i\leq N_{im}}$, the whole set of distortion coefficients.
\end{itemize}
Considering the assumptions made in Section \ref{sec:noise} on the noise statistical distribution, the Maximum Likelihood estimation of these unknowns boils down to the Weighted Least Square minimization of the following criterion (e.g. \cite{Mugnierbook}): \\
\begin{equation}
\label{eq:criterion}
\resizebox{\hsize}{!}{$
J\Big(\big[X^{ref}\big];\big[A~\big]\Big)=\left\{
\begin{array}{ll}
         \sum\limits_{j=1}^{N_{star}} \sum\limits_{i=1}^{N_{im}} w_{i,\,j}\left|X_{i,\,j}^{data}-\left(x^{ref}_{ j}+\sum\limits_{l=0}^{d}  \sum\limits_{k=0}^{l} a_{\,i,\,m,\,x}~{L_{l-k}(x_{j}^{ref})}{L_{k}(y_{j}^{ref})}\right)\right|^2 \\
        \sum\limits_{j=1}^{N_{star}}  \sum\limits_{i=1}^{N_{im}} w_{i,\,j}\left|y_{i,\,j}^{data}-\left(y^{ref}_{ j}+\sum\limits_{l=0}^{d}  \sum\limits_{k=0}^{l} a_{\,i,\,m,\,y}~{L_{l-k}(x_{j}^{ref})}{L_{k}(y_{j}^{ref})}\right)\right|^2 \\
\end{array}
\right.$}
\end{equation}
\noindent where $w_{i,\,j}$ is a weighing coefficient that can be used to quadratically weight the impact of each data in the criterion according to their Signal to Noise Ratio (SNR): 
\begin{equation}
w_{i,\,j}=\frac{1}{\sigma_{meas,\,i,\,j}^2}
\end{equation}
~\\
\noindent Note that the number of available reference sources governs the maximum order of coordinate transformations: A 1st order polynomial transform with six parameters requires at least three reference sources, i.e., six coordinates. A 2nd-order transform (12 parameters) requires at least six reference sources, and so on. In short, we must have $N_{star} \geq N_\text{modes}$.

\subsection{Practical implementation}
This section details the practical implementation for the minimization of the criterion defined in Equation (\ref{eq:criterion}).
As mentioned previously, the aim of the algorithm is to perform a joint estimation of both:
\begin{itemize}
\item[$\cdot$]  the reference positions of all the stars $\big[X^{ref}_{j}\big]_{1\leq j\leq N_{star}}$ noted $\big[X^{ref}\big]$; 
\item[$\cdot$]  the distortion coefficients of all frames $\big[A_{i}\big]_{1\leq i\leq N_{im}}$ noted $\big[A~\big]$
\end{itemize}
In practice, this criterion is minimized in an alternate fashion with respect to one set of variables, while the second is set. We detail hereafter the two steps included in one iteration of the minimization. For convenience, we detail the equation corresponding to the $x$-coordinates only, but the expressions can transposed to the $y$-coordinates. The estimation of a parameter $p$ is noted $\widehat{p}$.\\

%\noindent First step: Estimation of the reference positions $[X^{ref}]$ with $\big[A~\big]$ set. %In this case, we can define a sub-criterion, $J_1$:
The first step of the alternate estimation is to find $[X^{ref}]$ for the current value of $\big[A~\big]$. For a given $\big[A~\big]$, it is easy to realize that criterion  $J\Big(\big[X^{ref}\big];\big[A~\big]\Big)$ of Equation (\ref{eq:criterion}) can be written as a sum of $N_{star}$ independent terms:
\begin{eqnarray}
J\Big(\big[X^{ref}\big];\big[A~\big]\Big)=
        % \sum\limits_{j=1}^{N_{star}} \sum\limits_{i=1}^{N_{im}} w_{i,\,j}\left|X_{i,\,j}^{data}-\left(x^{ref}_{ j}+\sum\limits_{l=0}^{d}  \sum\limits_{k=0}^{l} a_{\,i,\,m,\,x}~{L_{l-k}(x_{j}^{ref})}{L_{k}(y_{j}^{ref})}\right)\right|^2 \\
         \sum\limits_{j=1}^{N_{star}}  J_1\Big(X^{ref}_j,\big[A~\big]\Big)%$}
\end{eqnarray}
where:
\begin{eqnarray}
\resizebox{\hsize}{!}{$
J_1\Big(X^{ref}_j;\big[A~\big]\Big)=\sum\limits_{i=1}^{N_{im}} w_{i,\,j}\left|X_{i,\,j}^{data}-\left(x^{ref}_{ j}+\sum\limits_{l=0}^{d}  \sum\limits_{k=0}^{l} a_{\,i,\,m,\,x}~{L_{l-k}(x_{j}^{ref})}{L_{k}(y_{j}^{ref})}\right)\right|^2$}
\end{eqnarray}
Hence, the estimation of the reference positions can be done independently for each star $j$ and is obtained as:
\begin{equation}
{\widehat{X}_j}^{ref}= \arg \underset{X_j^{ref}}{\min} \left[ J_1\left(X_j^{ref};[A\,]^{}\right)\right]
\end{equation}
%can be written as a sum of $N_{im}$ 
The same reasoning can be applied for the second step. For a given $\big[X^{ref}\big]$, criterion  $J\Big(\big[X^{ref}\big];\big[A~\big]\Big)$ of Equation (\ref{eq:criterion}) can be written as a sum of $N_{im}$ independent terms:
%\noindent Estimation of distortion coefficients $\big[A~\big]$ with $[X^{ref}]$  set:
\begin{eqnarray}
J\Big(\big[X^{ref}\big];\big[A~\big]\Big)=\sum\limits_{i=1}^{N_{im}}  J_2\Big(\big[X^{ref}\big],A_i\Big)%$}
\end{eqnarray}
where: 
\begin{eqnarray}
\resizebox{\hsize}{!}{$
J_2\Big(A_i~;\big[X^{ref}\big]\Big) = \sum\limits_{j=1}^{N_{star}} w_{i,\,j}\left|X_{i,\,j}^{data}-\left(x^{ref}_{ j}+\sum\limits_{l=0}^{d}  \sum\limits_{k=0}^{l} a_{\,i,\,m,\,x}~{L_{l-k}(x_{j}^{ref})}{L_{k}(y_{j}^{ref})}\right)\right|^2 $}
\end{eqnarray}
The estimation of the distortion coefficients can thus be done independently for each frame $i$:
\begin{equation}
{\widehat{A}_i}^{}= \arg \underset{A_i}{\min} \left[ J_2\left(A_i~;\big[X^{ref}\big]^{}\right)\right]
\end{equation}
These two steps are performed sequentially until a stopping condition relative to the convergence is reached. For exemple, we choose to stop the convergence when the difference of the estimates between two iterations is lower than $10^{-5}$:
\begin{equation}
C_{A}^{(k)}=\frac{\left| {\big[\widehat{A}~\big]}^{(k-1)} -{\big[\widehat{A}~\big]}^{(k)} \right| }{{\big[\widehat{A}~\big]}^{(k)} } \leq 10^{-5}
\end{equation}
\noindent and,
\begin{equation}
C_{X^{ref}}^{(k)}=\frac{\left| {\big[\widehat{X}^{ref}~\big]}^{(k-1)} -{\big[\widehat{X}^{ref}~\big]}^{(k)} \right| }{{\big[\widehat{X}^{ref}~\big]}^{(k)} } \leq 10^{-5}
\end{equation}
where 
$\widehat{p}^{~(k)}$ is the estimation of parameter $p$ at the $k$-th iteration.
To summarize, a bloc diagram of the algorithm is presented hereafter (Figure \ref{fig:algo}).
%The \textit{correction step} is described is Section \ref{sec:estpara}.

% Define block styles
\tikzstyle{decision} = [diamond, draw, %fill=blue!20, 
    text width=7.em, text centered, node distance=3cm, inner sep=0pt]
\tikzstyle{blocks} = [rectangle split,rectangle split parts=2, draw, %fill=blue!20, 
    text width=8em, text centered, rounded corners, minimum height=4em]

\tikzstyle{block4} = [rectangle split,rectangle split parts=2, draw=white, %fill=blue!20, 
    text width=10.2em, text centered, rounded corners, minimum height=7em]

\tikzstyle{block} = [rectangle, draw, %fill=blue!20, 
    text width=6em, text centered, rounded corners, minimum height=4em]
    \tikzstyle{block2} = [rectangle split,rectangle split parts=2, draw,% fill=blue!20, 
    text width=20em, text centered, rounded corners, minimum height=7em]

    \tikzstyle{block3} = [rectangle split,rectangle split parts=2, draw,% fill=blue!20, 
    text width=15em, text centered, rounded corners, minimum height=7em]

\tikzstyle{line} = [draw, -latex']
\tikzstyle{cloud} = [draw, rectangle,fill=red!20, node distance=3cm,
    minimum height=2em]

\tikzstyle{box}=[draw, minimum width=5em, 
    text centered, rounded corners,minimum height=2.5em] % draw=white

\tikzset{
    pics/vhsplit/.style n args = {3}{
        code = {
        \node[text width=2cm] (out) at (0,-16.85) {#1};  
        \node[anchor=north east,text width=2cm] (B) at (out.south) {#2};
        \node[anchor=north west,text width=2cm] (C) at (out.south){#3};
        \node[inner sep=0pt,draw,rounded corners,fit=(out)(B)(C)] {}; 
        \draw (B.north west) -- (C.north east)
              (B.south east) -- (C.north west);    
        }
    }
}

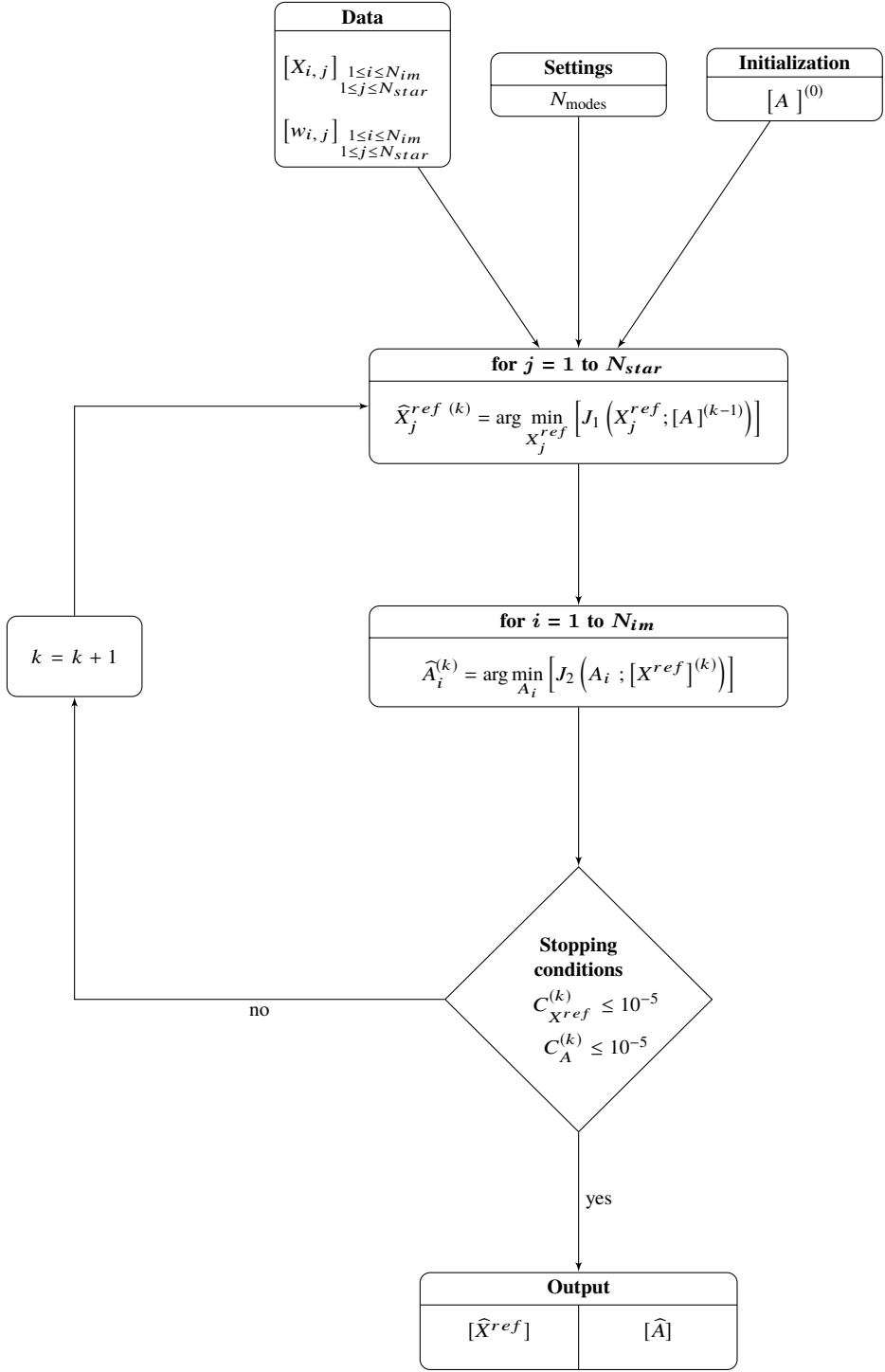
\begin{figure*}
\vspace{1cm}	
\begin{tikzpicture}[node distance = 3cm, column  sep=1cm,row  sep=2cm, align=center,auto]
    % Place nodes
    \node [blocks] (set) {\textbf{Settings} \nodepart{second}$N_\text{modes}$};

    \node [blocks, left of=set] (data) {\textbf{Data} \nodepart{second} \begin{equation*} \left[X_{i,\,j}\right]_{\substack{1 \leq i \leq N_{im}\\1 \leq j \leq N_{star}}} \end{equation*} \begin{equation*} \left[w_{i,\,j}\right]_{\substack{1 \leq i \leq N_{im}\\1 \leq j \leq N_{star}}} \end{equation*}};
   
    \node [blocks, right of=set] (init) {\textbf{Initialization}\nodepart{second} $\big[A~\big]^{(0)}$};
   
    \node [block2, below of=set, node distance=4.5cm] (Xref) {\textbf{for \boldsymbol{$j$} \boldsymbol{$=$} \boldsymbol{$1$} to \boldsymbol{$N_{star}$} } \nodepart{second} $$
{\widehat{X}_j}^{ref~(k)}= \arg \underset{X_j^{ref}}{\min} \left[ J_1\left(X_j^{ref};[A\,]^{(k-1)}\right)\right]
$$};
   
    \node [block2, below of=Xref, node distance=3.5cm] (A) {\textbf{for \boldsymbol{$i$} \boldsymbol{$=$} \boldsymbol{$1$} to \boldsymbol{$N_{im}$}} \nodepart{second} $$
{ \widehat{A}_i}^{(k)}= \arg \underset{A_i}{\min} \left[ J_2\left(A_i~;\big[X^{ref}\big]^{(k)}\right)\right]
$$};
   
    \node [block, left of=A, node distance=7cm] (k) {$k=k+1$};
    \node [decision, below of=A,node distance=4.8cm] (decide) {\textbf{Stopping conditions}
 \begin{eqnarray*}&C_{X^{ref}}^{(k)} \leq 10^{-5}\\
& C_{A}^{(k)}\leq 10^{-5} \end{eqnarray*}};

%    \node [block4, below of=decide, node distance=5cm] (stop) {\textbf{Output} \nodepart{second} \\\hspace{-0.4cm} $ [\widehat{X}^{ref}]  \hspace{0.52cm} \Bigg| \hspace{0.8cm} [\widehat{A}]$~\\};

    \node [block4, below of=decide, node distance=4.8cm] (stop) {\vspace{0.56cm} \nodepart{second}\vspace{1cm}};
    
      \path pic (out) {vhsplit={\textbf{Output}}{~\\$ [\widehat{X}^{ref}] $\\~}{~\\ $[\widehat{A}]$\\~}};

%\node [block3, below of=stop,right of=stop, node distance=3.5cm] (corr) {\textbf{for \boldsymbol{$j$} \boldsymbol{$=$} \boldsymbol{$1$} to \boldsymbol{$N_{star}$} \\ for \boldsymbol{$i$} \boldsymbol{$=$} \boldsymbol{$1$} to \boldsymbol{$N_{im}$}} \nodepart{second} 
%$${\widehat{X}_{i,\,j}}^{corr}= arg \underset{X_{i,\,j}^{corr}}{min} \left[ J_3\left(X_{i,\,j}^{corr},[\widehat{A}\,]\right)\right]
%$$};
    % Draw edges
    \path [line] (set) -- (Xref);
    \path [line] (Xref) -- (A);
    \path [line] (A) -- (decide);
    \path [line] (decide) -| node [near start] {no} (k);
    \path [line] (k) |- (Xref);
    \path [line] (decide) -- node {yes}(stop);
    \path [line] (data) -- (Xref);
    \path [line] (init) -- (Xref);
   % \path [line] (stop) -- node {correction}(corr);
%\path (stop.second east) edge node [above, sloped, ->] {correction} (corr);
%\path [line](stop.second east)  -| node [near end]  {\textit{correction} \textit{step}} (corr);
\end{tikzpicture}
\caption{Bloc diagram of the minimization algorithm of Equation (\ref{eq:criterion}) multi-variable $J$ criterion.}
\label{fig:algo}
\end{figure*}

When both parameters, $[\widehat{A}]$ and $[\widehat{X}^{ref}]$ are estimated, both of them can be useful and their use depends on the final physical objective. The reader's attention is here drawn to one important point: the parameters are estimated regarding the data set provided.
If the instrument suffers from static distortions, for exemple a global shift, then the data collected with this instrument are all globally shifted regarding to the celestial coordinates. The same holds true if the instrument suffers from a global scale factor: the data collected with this instrument are all globally scaled.
These global modes cannot be measured unless
% a large set of on-sky observations with different offsets and orientations is available, then, the method get to the self-calibration method describes by \cite{Anderson2003}, or  
a priori knowledge of the static distortion is available from a previous calibration. In the latter case, the proposed method derives distortion solutions calibrated for each observation set and for each frame, that varies around the static distortion provided. In the following, we address this configuration and consider a priori knowledge of the instrument's static distortion. In practice, the  static distortion is implemented as initial parameters for the distortion coefficients $[A\,]^{0}$.

\subsection{Estimated parameters}
\label{sec:estpara}
Both estimated parameters $[\widehat{A}]$ and $[\widehat{X}^{ref}]$, can be used independently depending on the scientific aim of the study at hand.\\
In the case of absolute astrometry studies, the information required is simply the estimate of the reference positions $[\widehat{X}^{ref}]$, obtained as described above.

%For classical observation, if we want the estimated reference positions to match with celestial coordinates, then the static distortion is needed to derive the true positions of the stars. 
%In the case a large set of on-sky observations with different offsets and orientations is available, all the distortion modes can 
%the first one is that a good estimation of the distortion-free positions, $[\widehat{X}^{ref}]$ require a . This point is detailed in Section \ref{sec:nim}. The second is tha
%and the proposed method cannot access this information  data alone.

%The answer to that problem is given by the setting of the starting point of the minimization: by using the value of the global modes as starting point, we make sure that they are taken into account in the estimation and that we converge through the right solution.
%In practice, these global modes correspond to the average distortion expected in the data, i.e. the static part of the distortions. These might be calibrated by external mesures as previously mentioned in Sec \ref{sec:2}.

\noindent In the case of relative astrometry or photometry studies, an additional step is needed, namely to \textit{correct} the different frames from the distortion. This, aiming to place stellar positions in a relative global reference frame or to stack the images. % (e.g. \textit{correction step} in the algorithm diagram). 
In that case, the parameters of interest are the estimated distortion coefficients $[\widehat{A}\,]$, of all frames. The correction of the distortion may apply to the star positons only, or to the entire image thanks to an interpolation process. In the latter case we want to correct star positions only, the corrected position of each star $j$, in the frame $i$ is noted ${X_{i,\,j}}^{corr}$ and calculated independently as follow:

\begin{equation}
{\widehat{X}_{i,\,j}}^{corr}= \arg \underset{X_{i,\,j}^{corr}}{\min} \left[ J_3\left(X_{i,\,j}^{corr};[\widehat{A}\,]\right)\right]
\end{equation}

\noindent where :
\begin{equation}
\resizebox{\hsize}{!}{$
J_3\left(X_{i,\,j}^{corr};[\widehat{A}\,]\right)= \left|X_{i,\,j}^{data}-\left(x^{corr}_{i,\,j}+\sum\limits_{l=0}^{d}  \sum\limits_{k=0}^{l} a_{\,i,\,m,\,x}~{L_{l-k}(x^{corr}_{i,\,j})}{L_{k}(y^{corr}_{i,\,j})}\right)\right|^2  $}
\end{equation}
\noindent and equivalent for the $y$-coordinates.\\

Note that $J_1$ is the sum of $N_{im}$ $J_3$-like terms, so that ${\widehat{X}_{i,\,j}}^{corr}$ is the reference position of star $j$ as predicted by the distortion map $\widehat{A}_i$ alone. Whereas $\widehat{X}_{j}^{ref}$ is the reference position of star $j$ that best fits all the estimated distortion maps.\\

\noindent The effect of the algorithm on one particular star $j_0$, is illustrated by Figure \ref{Figure:schema}: the left panel shows the set of $\left[X^{data}_{i,\,j_0}\right]_{1 \leq i \leq Nim}$ before distortion correction (black crosses), as well as the true reference position $X^{ref}_{j_0}$ (red cross). The right panel shows the set of positions corrected from distortion $\left[\widehat{X}^{corr}_{i,\,j_0}\right]_{1 \leq i \leq Nim}$ (black crosses), using the estimated distortion coefficient $[\widehat{A}\,]$.  Bottom panel is a zoom of right panel: it additionally shows  
the estimated reference position ${\widehat{X}^{ref}}_{j_0}$ (blue cross) and the averaged corrected positions on all images (orange cross).  Note that these two positions match each other: $avg_{i\,}(\widehat{X}^{corr}_{i,\,j_0})\simeq\widehat{X}^{ref}_{j_0}$.  
This result is expected from the definition of the $J_1$ and $J_3$ criteria and it can easily be demonstrated in the case of linear transformations (i.e. $d=0$). In practice, we observe that this equality is verified for any transformation order, within a few $10^{-3}$ pixels.
The algorithm performance can then be quantified using two error criteria illustrated on the bottom panel : a bias
%, calculated as the error of estimation on the reference position $\widehat{X}^{ref}_{j_0}$, noted $b_{j_{0}}$ and 
represented as a blue arrow; and a standard deviation
%on the corrected positions $\left[\widehat{X}^{corr}_{i,\,j_0}\right]_{1 \leq i \leq Nim}$, noted $\sigma_{j_{0}}$ and 
shown as a black arrow. Both error criteria are described analytically in the next section (\ref{sec:error}).

\subsection{Error criterion}
\label{sec:error}

\begin{figure}
\includegraphics[width=\hsize]{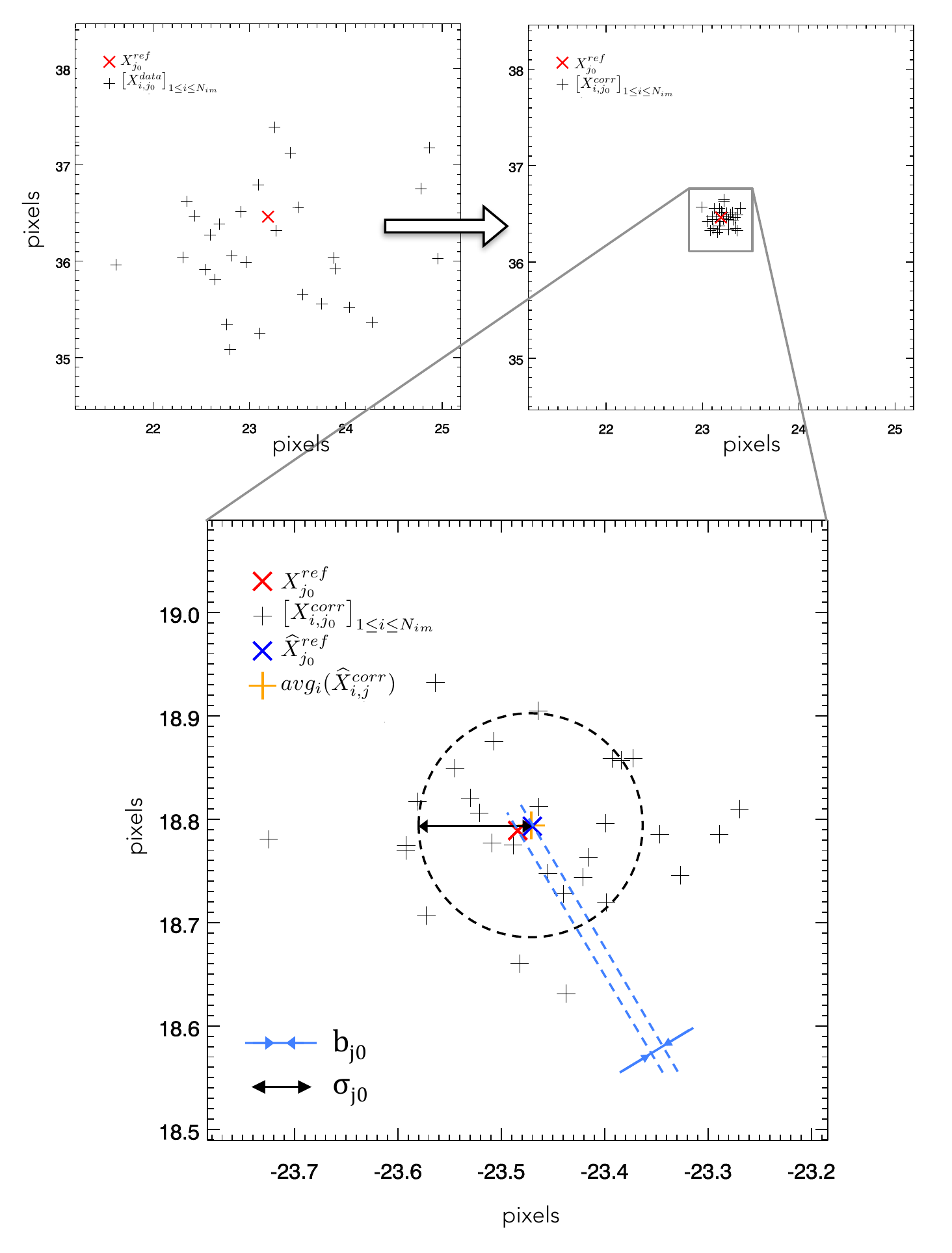}
\caption{Illustration of the algorithm effect on the star $j_0$, in the case of $Nim=30$. Left panel: set of $\left[X^{data}_{i,\,j_0}\right]_{1 \leq i \leq Nim}$ before distortion correction (black crosses) and true reference position $X^{ref}_{j_0}$ (red cross). Right panel: set of positions corrected from distortion $\left[\widehat{X}^{corr}_{i,\,j_0}\right]_{1 \leq i \leq Nim}$ (black crosses). Bottom panel: zoom of right panel. It additionally shows  
the estimated reference position ${\widehat{X}^{ref}}_{j_0}$ (blue cross) and the averaged corrected positions on all images (orange cross).  Note that $avg_{i\,}(\widehat{X}^{corr}_{i,\,j_0})\simeq\widehat{X}^{ref}_{j_0}$. The two error criteria used to quantify the algorithm performance are also illustrated on bottom panel.
The first one is a bias: the distance between the true reference position,  $X^{ref}_{j_0}$ (red cross) and the estimated reference position, ${\widehat{X}^{ref}}_{j_0}$ (blue cross). It is noted $b_{j_{0}}$ and represented in blue arrow. The second one is the standard deviation of the corrected positions (schematized in black arrow) is noted $\sigma_{j_{0}}$. Both error criterion are described in further details in section \ref{sec:error}.}
\label{Figure:schema}
\end{figure}

As mentioned previously, the performance of the algorithm, is quantified by two criteria that we define as a bias and a standard deviation as follows:\\
\begin{itemize}
\item[$\cdot$]The bias quantifies the ability to estimate the reference positions. It is defined for each star by the distance between the estimated reference position, $\widehat{X}^{ref}_{j}$ and the reference position, $X^{ref}_{j}$. It is noted $b_j$ hereafter.
\begin{equation}
b_j= {\widehat{X}^{ref}}_{j}-X^{ref}_{j}
\end{equation}
This error is shown as a blue arrow in Figure \ref{Figure:schema} and can be quadratically averaged on all the stars:
\begin{equation}
b=\sqrt{\frac{1}{N_{star}}\sum_{j=1}^{N_{star}}\mid {\widehat{X}^{ref}}_{j}-X^{ref}_{j}\mid^2}\\
\end{equation}
~\\
\item[$\cdot$]The standard deviation, which quantifies the algorithm ability to place every stellar position in a global reference frame or to stack the images. It is the standard deviation of each star's corrected position through the stack of images and it is noted $\sigma_j$.
\begin{equation}
\sigma_j=\sqrt{\frac{1}{N_{im}}\sum_{i=1}^{N_{im}}\mid \widehat{X}_{i,\,j}^{corr}-avg_{\,i}(\widehat{X}^{corr}_{i,\,j})\mid^2}
\end{equation} 
\end{itemize}
This error is represented as a black arrow on Figure \ref{Figure:schema}. Note that this error is lower bounded by the measurement error $\sigma_{meas,~j}$ of each star $j$, which is the fundamental limit of the astrometric precision in the case of a perfect optical system:
\begin{equation}
\sigma_{meas,~j} \leq \sigma_j
\end{equation} 
\noindent The information given by this last criterion is very local in the field. It is calculated at the position of the reference sources used in the minimization.
In order to evaluate the ability to stack the whole image, this error criterion is generalized to stars that are not used in the resolution and that are distributed as a regular grid in the field. Hereafter, this generalized criterion is noted $\sigma$, the number of stars considered in the regular grid is noted $N_{star,~grid}$ and the stars included in the grid are indexed $j'$:

\begin{equation}
\sigma=\sqrt{ \frac{1}{N_{star,\,grid}}\sum_{j'=1}^{N_{star,\,grid}} \sigma_{j'}^2   }
\end{equation}

\section{Simulation results}
\label{sec:simu}

\begin{figure}
\includegraphics[width=\hsize,trim={0cm 0cm 0cm 0cm}]{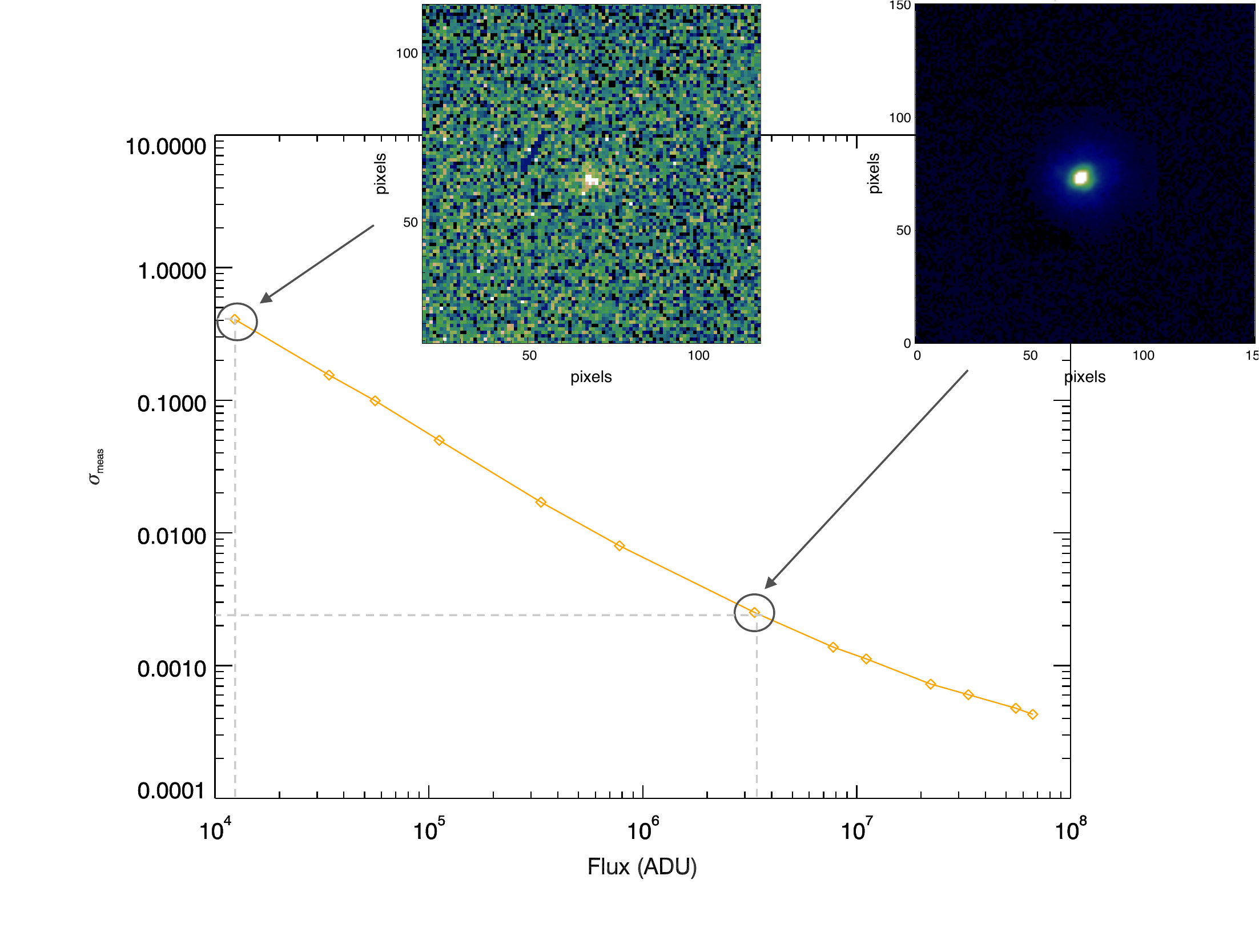}
\caption{Error on position measurements versus the flux on typical GeMS/GSAOI data. Are also represented, one star fitted with an error of $\sigma_{meas}=0.14$\,pixels on its positions and one star fitted with an error of $\sigma_{meas}=0.002$\,pixels on its positions, with a Levenberg-Marquart fitting tool. The last one corresponds to a 18 magnitude in \ks \,band for a 60\,s exposure. }
\label{Figure:star}
\end{figure}

%If no distortion is present in the data, or if distortions are perfectly corrected, the astrometric error is set by the error of position measurements mentioned earlier $\sigma_{meas}$.  
%If distortions are present in the data, a distortion error term is added to the astrometric error. %This is summarized by the expression : 
%\begin{equation}
%rms_{astro}=\sqrt{\sigma_{meas}^2+rms_{disto}^2}
%\end{equation}
%The algorithm described previously aim to minimize this distortion error term.
%If the static distortion of the instrument is perfectly known as well as the positions of the stars in the field (i.e., $\sigma_{meas}=0$), the algorithm performs a perfect correction. The corrected position of each star is then the exact reference position. In practice, this precision is never reached, due to several error terms that we detail in this section. These errors depend on: 
This section aims at validating the algorithm and investigating its performance using simulated data representative of those collected with the GeMS/GSAOI systems. The data are simulated according to the direct model described in Section \ref{sec:model} and the noise introduced in the position measurements is a Gaussian white noise (as explained in  Section \ref{sec:noise}), with a variance $\sigma^2_{meas}$. 
In typical  GeMS data, the error on position measurements ranged between 0.002\,pixels and 0.14\,pixels (0.02\,mas and 2.8\,mas respectively), mostly depending on the magnitude of the objects. For example, we show in Figure \ref{Figure:star} two stars, one fitted with an error of $\sigma_{meas}=0.002$\,pixel, and one fitted with an error of $\sigma_{meas}=0.14$\,pixel, in typical GeMS/GSAOI data with a Levenberg-Marquart  fitting tool. The last one represents the faintest stars detectable in a single frame, corresponding to a 18 magnitude in \ks \,band for a 60\,s exposure. \\

For each simulation, the performance are evaluated regarding the two error criteria defined in Section \ref{sec:error}. We recall here that an illustration of these two error criterion is given in Figure \ref{Figure:schema}. 
The first part of this section aims at validating validate the algorithm using one classical simulated data set: we observe here the influence of the reference sources positions. In a second part of this section, we investigate the noise propagation and the influence of observation's parameters such as the number of reference sources in the field or the number of frames available.
 In sections \ref{sec:valid} to \ref{sec:noiseprop}, the number of distortion modes introduced is equal to the number of distortion mode searched in the data : $N_\text{modes}=N_\text{modes,\,search}$ while in the section \ref{sec:alias} which is dedicated to the study of the aliasing effect, we consider $N_\text{modes} \geq N_\text{modes,\,search}$.
The settings used for each simulation are summarized in Table \ref{table:simu}.

%\begin{itemize}
%\item Intrinsic proprieties of the data: the level of noise in the data measurement $\sigma_{meas}$, the number of stars, $N_{star}$, and frames $N_{im}$
%\item The fine-tuning of the algorithm : the starting point of the minimization and the number of mode searched in the data, which both depend on a priori-knowledge of the instrument's static distortion. 
%\end{itemize}
%Then we will split the distortion error into two term : the estimation error term $\sigma_{est}$, which include the error on the estimation of both parameters (the distortion coefficient and the reference positions), and the correction error term, $\sigma_{corr}$, which include the correction of the position once the parameters are estimated. \subsection {Estimation of the distortion}

\subsection{Validation of the algorithm}
\label{sec:valid}
%In this section, we investigate the behavior of the algorithm confronted to noisy data, using simulations. The performance are evaluated regarding the two error criteria described in the previous section (\ref{sec:error}), the data are formed using the direct model described in Section \ref{sec:model} and 

This first simulation presented here aims to illustrate the validation the algorithm using one classical simulated data set. We consider here 30 reference sources in the field and 10 frames ($N_{star}=30$ and $N_{im}=10$). Ten distortions modes are introduced on both the $x$-axis and the $y$-axis (i.e.,  $N_\text{modes}=10$, and $d=3$) and the level of noise considered is $\sigma_{meas}=0.14$\,px (it corresponds to the faintest stars that can be detected with the GeMS/GSAOI system in a 60\,s exposure : \ks \,magnitude = 18). This simulation is referred to as \textit{Simulation 1 } in Table \ref{tab:simu}.\\
Using these parameters and this pessimist level of noise, the precision obtained on the estimation of the reference positions is $b=0.05$\,pixel which corresponds to 1\,mas on GeMS data.   \\
The standard deviation of the corrected position, $\sigma_{j'}$ is calculated on a of grid of $150 \times 150$ stars (i.e. $0 \leq j' \leq 150 \times 150)$ regularly distributed  in the image. The results are shown in Figure \ref{Figure:ergrid}. Black corresponds to an error of  $\sigma_{j'}=0.05$ pixels and white corresponds to a value of  $\sigma_{j'}=0.7$\,pixels.  On this Figure, white crosses show the positions of the reference sources used in the minimization. As one might expect, the performance dramatically drops on the border of the field as no reference-source is constraining the model.
The standard deviation of the corrected position averaged on the whole grid is $\sigma_\text{grid}= 0.2$ pixels, while the averaged value calculated inside the square containing most of the stars used in the minimization (i.e., inside the red contour) is $\sigma_\text{contour}=0.16$ pixels (dark blue).
In the following, the standard deviation, $\sigma$ is measured on a regular grid of $150 \times 150$ stars included in a given square containing most of the reference stars used in the minimization (i.e., in this particular simulation, inside the red contour represented in Figure \ref{Figure:ergrid}):

\begin{figure}
\includegraphics[width=\hsize,trim={0cm 0cm 0cm 0cm}]{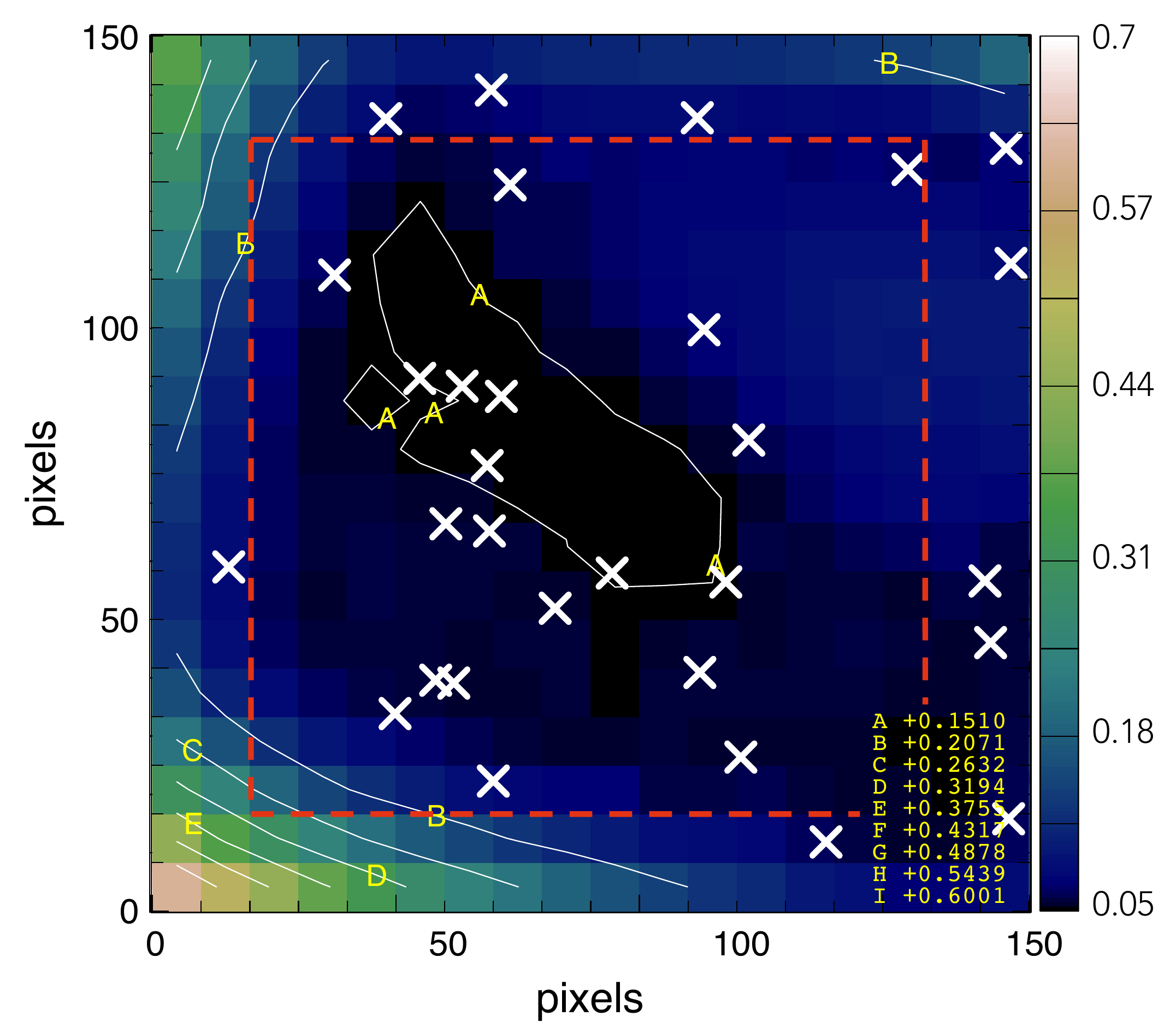}
\label{Figure:ergrid}
\caption{Distribution of the error $\sigma_{j'}$ in the field. The error is calculated on a regular $150 \times 150 $ grid of star. The white crosses represent the positions of the reference stars used in the minimization, $\big[X^{ref}\big]_{1\leq j \leq N_{star}}$. Max is white ($\sim 0.7$\,pixels) and min is black ($\sim 0.05$\,pixels). The averaged value inside the red contour) is $\sigma_\text{contour}= 0.16$\,pixels (dark blue). Simulation runs with $Nim=10$, $N_{star}=30$, and $\sigma_{meas}=0.14$\,pixels.}
\end{figure}

\subsection{Noise propagation}
\label{sec:noiseprop}
\begin{figure}
\includegraphics[width=\hsize,trim={0cm 0cm 0cm 0cm}]{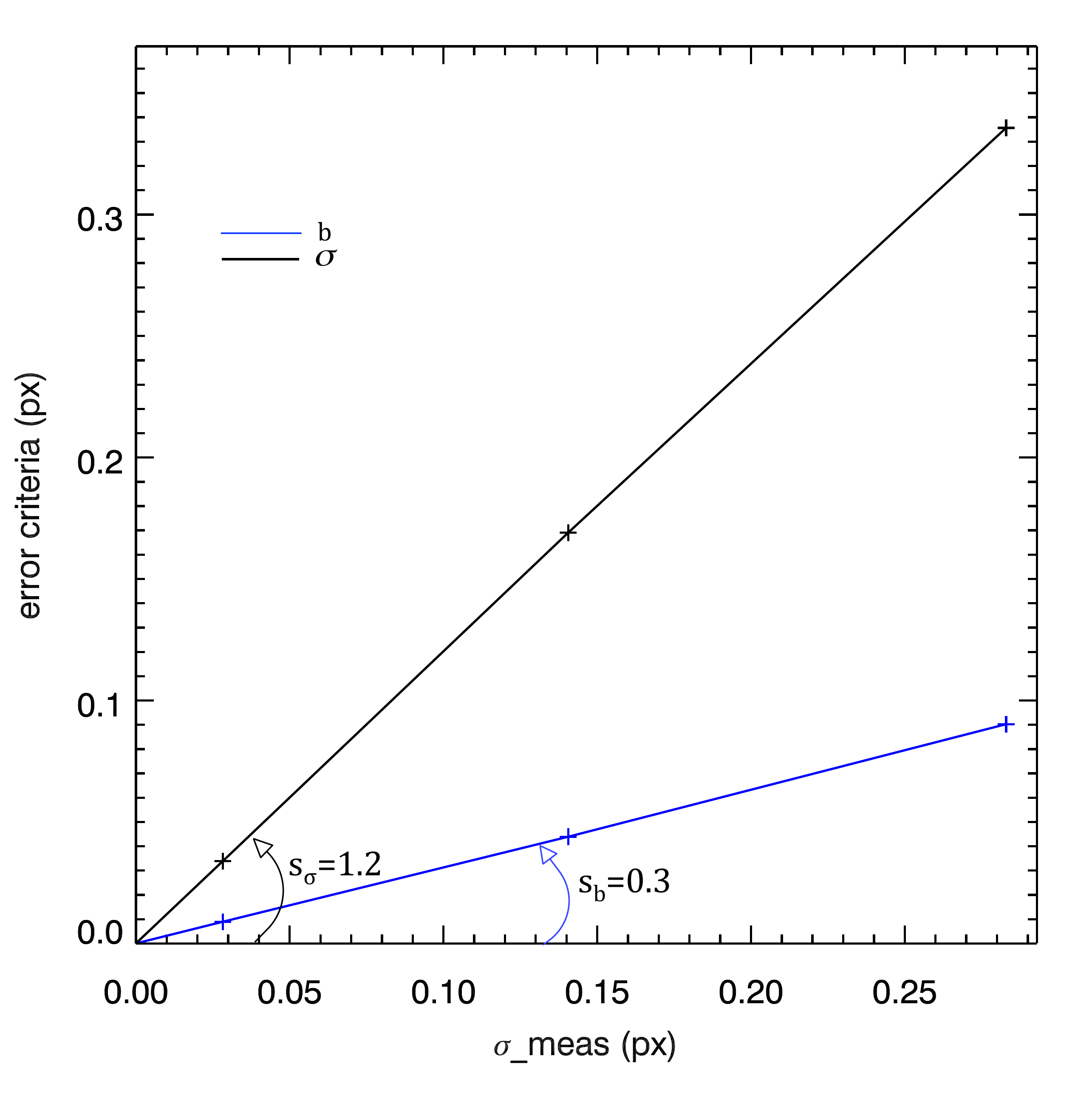}
\caption{Noise propagation on the two error criterion. In black, $\sigma$ is the standard deviation of the corrected positions. In blue, $b$ is the error on the estimation of the reference. Both are shown versus the standard deviation of the noise in the star position measurements, $\sigma_{meas}$. For each level of noise, $n=100$ outcomes are computed, the error criterion are then quadratically averaged on the $n$ outcomes. The slopes of the noise propagation curves are noted $s_{\sigma}$ and $s_{b}$, respectively for the standard deviation of the corrected positions slope and the error on the reference estimation slope. Simulation runs with $Nim=10$ and  $N_{star}=30$. For $\sigma_{meas}=0.14$\,pixels, we found back the results of Section \ref{sec:valid}: $\sigma=0.16$\,pixels and $b=0.05$\,pixels.}
\label{Figure:erbilan}
\end{figure}

We now assess the noise propagation in the minimization using the same configuration: $Nb_{star}=30$ and $N_{im}=10$. 
Simulations are carried out for different levels of noise :\\$\sigma_{meas} \in $[0,0.03,0.07,0.14,0.28]\,pixels, and for each level of noise $n=100$ outcomes are computed. The two error criteria described previously are quadratically averaged on the $n$ outcomes. This study is referred as \textit{Simulation 2} in Table \ref{tab:simu}.
The results is shown in Figure \ref{Figure:erbilan}: the error on the estimation of the reference position, $b$ is plotted in blue line and the standard deviation of the corrected position $\sigma$, is plotted in black line. Both are plotted versus the standard deviation of the noise in the positions measurement of the data, $\sigma_{meas}$. \\
%To get physical informations on the distortion map estimation quality, we introduced here two other error criterion : $er\_map$ and $er\_map,grid$, which converts the error on the coefficients into errors on the distortion map estimation (in pixels). The $er\_map$ give an idea of the error locally in the field, at the position of the stars used for the algorithm. The $er\_map,grid$ give the error on the distortion map estimation on a regular grid in the fiel. Of course, this error depends on the number of points included in that regular grid. We chose a large number of points with which, the error was no longer evolving. As we could expect, the error on the estimation of the distortion map is lower at the position of the stars used in the algorithm. In the following, we consider $er\_map,grid$, as we are evaluating the estimation of the correction in the whole image, and not only in local part of the field. %These errors can be interpreted as a variance and a biais as follow: 
%To go further, considering one single level of noise (for example $\sigma_{meas}=0.15 $ px), we can look at the $er\_photo_j$ error's evolution in the field. 
%The Figure \ref{Figure:erphotogrid} show the error $er\_photo_j$ calculated on a $300 \times 300 $ grid regularly distributed in the field. 
We find again the results derive for $\sigma_{meas}=0.14$\,pixels (pessimist case) in the previous section and observe that 
the noise propagation on the two error criteria follows a linear comportement which slopes are noted $s_{b}$ and $s_{\sigma}$ hereafter (respectively for the slope of the noise propagation on the reference position estimation and for the slope of the noise propagation on the corrected position standard deviation). These slopes correspond to the amplification of the noise, which depends on intrinsic properties of the observation set, such as the number of reference sources and the number of frames available and used in the minimization. We detail both influences in the next sections. \\

%\begin{equation}
%rms_{disto}=\sqrt{er\_Xref^2 + er\_map,grid^2}
%\end{equation} 
%\noindent With $er\_Xref$, the bias of estimation and $er\_map,grid$, the standard deviation.

%\begin{equation}
%rms_{disto}=er\_map,grid=\sqrt{er\_Xref^2+variance}
%\end{equation} 
%\noindent With $er\_Xref$, the bias of estimation.\\
 
% Les erreurs sont d\'ecorr\'el\'ees. A partir d'ici deux utilisations possible: 
%\begin{itemize}
%\item on a beaucoup d'\'etoiles, et on cherche les positions vraies: Dans ce cas l\`a, on utilise directement les Xref\_est.
%\item on cherche \`a corriger la distortion sur les images: dans ce cas l\`a, on utilise les coefficients estim\'s. et $rms_{disto}=er\_map,grid$
%\end{itemize}

%The error of estimation is define as :
%\begin{equation}
%\sigma^2_{est}=er\_Xref^2+er\_map,grid^2
%\label{eq:err}
%\end{equation}

%\begin{figure}
%\includegraphics[width=\columnwidth,trim={6cm 5cm 6cm 5cm}]{fig/er_A_norm_N_{im}10N_{star}20_n100-deg3_nocorrection_initA=avg_note2802.pdf}
%\caption{Propagation du bruit sur les erreur d'estimation des coefficients, $er\_A\_m$ pour $0 \leq m \leq nb\_mode -1$. La Figure de gauche represente l'erreur brut sur chaque coefficient, la Figure de droite represente l'erreur normalisee par l'amplitude moyenne de chaque coefficient.}
%\label{Figure:erA}
%\end{figure}

\subsubsection {Impact of the number of frames}
\label{sec:nim}
\begin{figure}
\includegraphics[width=\hsize,trim={0cm 0cm 0cm 0cm}]{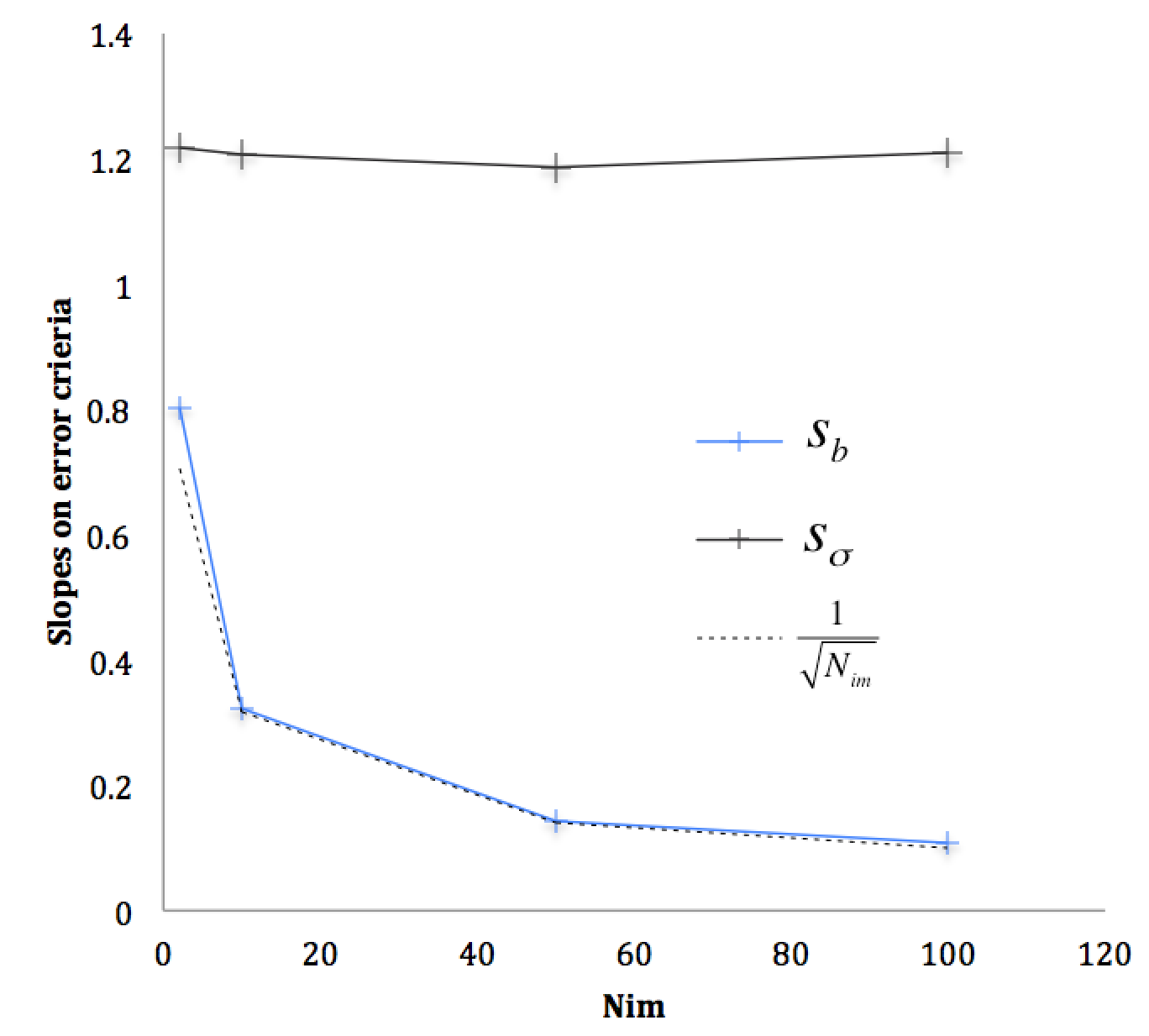}
\caption{Influence of the number of frame. Black curve is the standard deviation of the corrected positions slope ($s_{\sigma}$), and blue curve is the estimation of the reference error slope ($s_b$). The curve $y=1/\sqrt{x}$ is shown in grey dotted line. Simulation runs with $N_{star}=30$.
For $Nim= 10$, we found back the results of Section \ref{sec:noiseprop}: $s_{\sigma}=1.2$\,pixels and $s_b=0.3$\,pixels.}
\label{Figure:erim}
\end{figure}

In this section, we investigate the influence of the number of frames available on the noise propagation. 
Simulations as described previously are run with $N_{star}=30$ and a varying number of frames $N_{im} \in [2,10,50,100]$. Then, the same error criteria are calculated (simulations referred as \textit{Simulation 3} in Table \ref{tab:simu}). The noise propagation still follows a linear behavior but this time, we are interesting in the slope of this linear propagation. Figure \ref{Figure:erim} shows the evolution of the error propagation slope of both error criteria, as the number of frames increases. The black curve shows the noise propagation on the corrected positions standard deviation slope, $s_{\sigma}$ as function of the number of frames. 
%The equation is:\\
%$$y_1=s_{\sigma}(N_{im})$$\\%=\frac{\sigma(N_{im})}{\sigma_{meas}}$\\
and blue curve shows the noise propagation on the reference position estimation slope, $s_{b}$ versus the number of frames.
%The equation is:\\
%$$y_2=s_{b}(N_{im})$$\\%=\frac{b(N_{im})}{\sigma_{meas}}$\\
The curves show that the standard deviation of the corrected positions is independent of the number of frames. On the contrary, the error on the reference position estimation decreases when increases the number of frames. More precisely, we find that within a good approximation,  the estimation of the reference precision improves as the square roots  $N_{im}$:
\begin{equation}
b\sim \frac{\sigma_{meas}}{\sqrt{N_{im}}}
\end{equation}
This is an expected behavior, as increasing the number of frames, increases the number of independent measurements while keeping the number of unknowns (star's reference positions) constant. 

\subsubsection {Impact of the number of reference sources}
\begin{figure}
\includegraphics[width=\hsize]{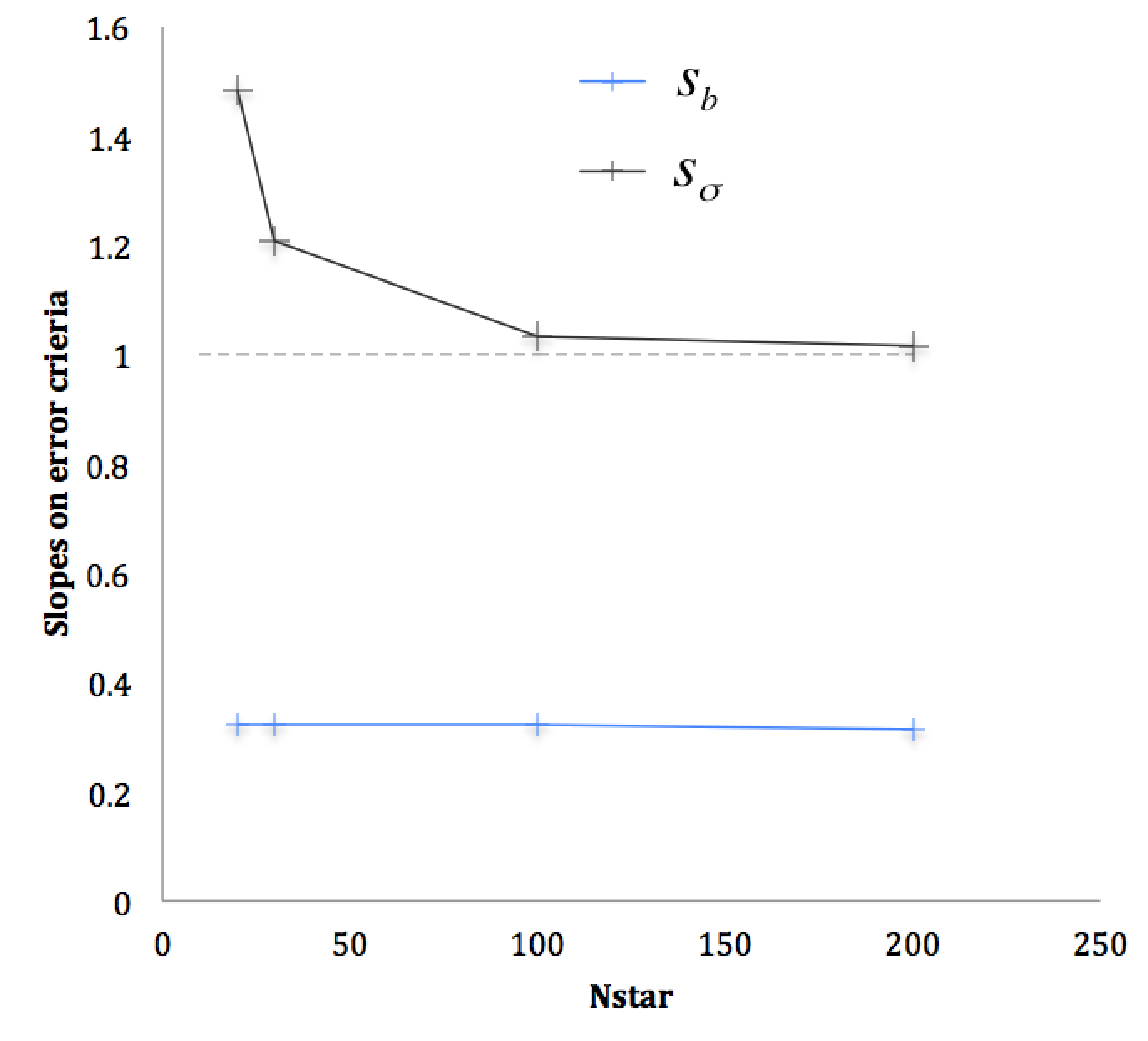}
\caption{Influence of the number of reference sources used in the minimization. Black curve is the slope on the standard deviation of the corrected positions ($s_{\sigma}$), and blue curve is the slope on the estimation of the reference error ($s_b$). The curve $y=1$ is shown in grey dotted line. Simulation runs with $Nim=10$. For $N_{star}= 30$, we found back the results of Sections \ref{sec:noiseprop} and \ref{sec:nim}: $s_{\sigma}=1.2$\,pixels and $s_b=0.3$\,pixels.}
\label{Figure:erstar}
\end{figure}

In this section we examine the influence of the number of reference sources used in the minimization following the same process as in the previous section. This time, the number of frame is fixed to $N_{im}=10$ while the number of star is varying: $N_{star} \in [20,30,100,200]$. This simulation is referred as \textit{Simulation 4} in Table \ref{tab:simu}. The results plotted in Figure \ref{Figure:erstar} show the slopes of the noise propagation on the two criteria versus the number of stars. The black curve shows the noise propagation on the corrected position's standard deviation slope $s_{\sigma}$, and the blue curve shows the noise propagation on the reference estimation slope $s_{b}$.
Contrary to the number of frames, the number of reference sources affects only the standard deviation of the corrected frame $\sigma$, while the estimation of the reference positions error, $b$ is not affected. This resulted from the fact that a higher number of reference sources allows a better estimation of the distortion in the whole field. 
In practice, with a hundred of reference sources available, the fundamental astrometric limitation mentioned in Section \ref{sec:error} is reached: $\sigma \simeq \sigma_{meas}$ which corresponds to $s_{\sigma} \simeq 1$.\\

 %Note that we find again the result giving in Section \ref{sec:error}:\\ $\sigma_{meas,~j} \leq \sigma_j$, with $s_{\sigma} \geq 1$.\\

In short, these results can be interpreted in the following way: the good estimation of a star position requires a large number of frames, i.e., a large number of measurements of this star position in the data. On the contrary, the ability to place every star in a same relative reference, or in other words, the ability to stack the images, is only governed by the number of reference sources available in the image to correctly estimate the distortion in the whole filed.   
%
%\begin{equation}
%er\_map,grid \sim \frac{1}{\sqrt{N_{star}}}
%\end{equation}

\subsection {Aliasing effects}
\label{sec:alias}
\begin{figure}
\includegraphics[width=\hsize]{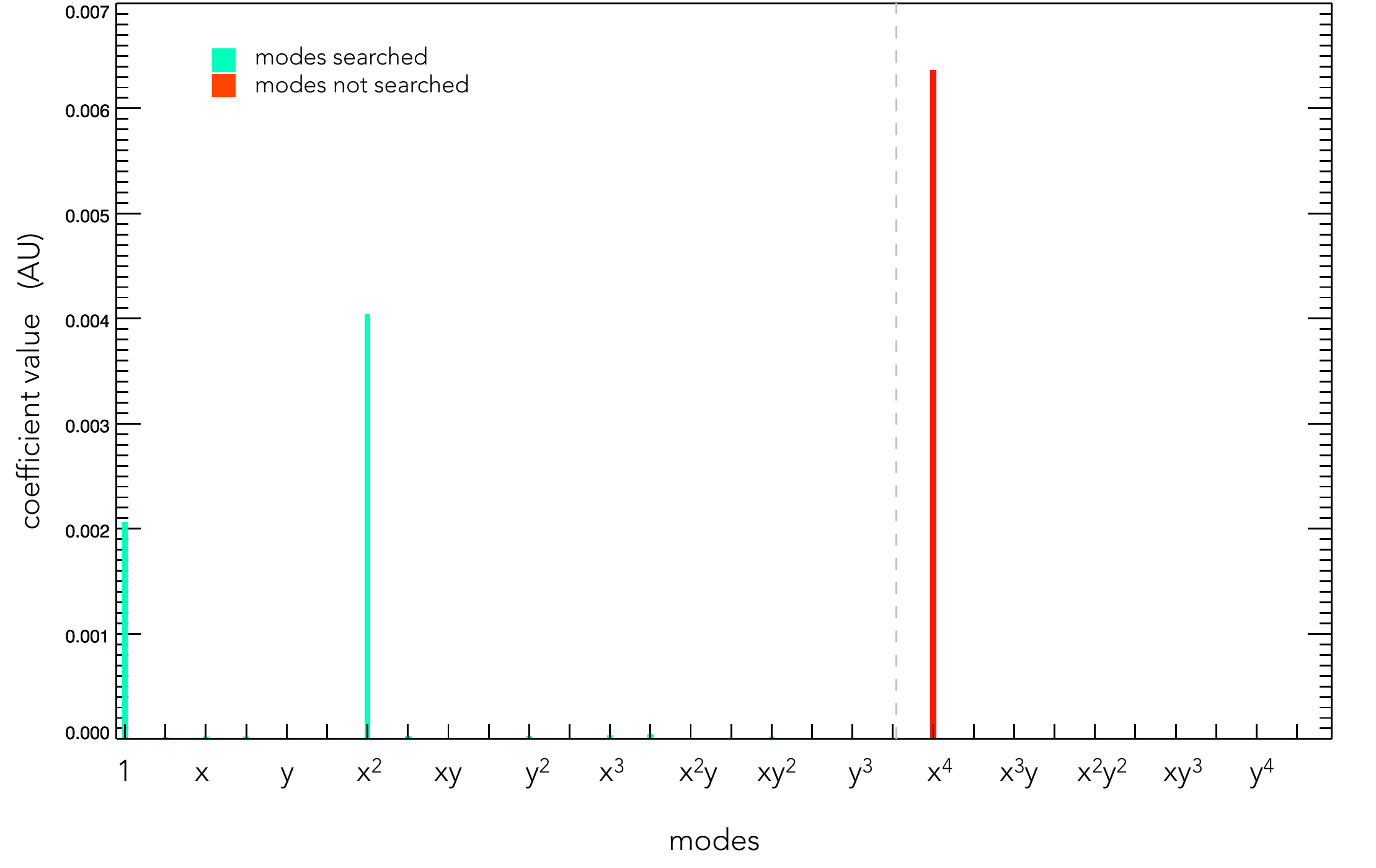}
\caption{Aliasing of the $m=11$ mode proportional to $x^4$ on the $3^{rd}$ and lowers order modes. Green bars represent the errors on the estimated coefficients. Red bar represent the value of the additional non-estimated coefficient. For each mode, the first value corresponds to the $x$-axis distortion and the second value corresponds to the $y$-axis distortion. Each mode is identified according to their order in $x$ and $y$ variables. 
}
\label{Figure:spec1}
\end{figure}

%We now try to estimate the aliasing error term noted $\sigma_{alias}$. 
In the previous simulations we supposed: $N_\text{modes} = N_\text{modes,\,search}$.
As aliasing effect will appear if the real number of distortion modes present in the data is higher than the number of modes taken into account in the model. When this happens, the non estimated modes are affecting the estimation of the other modes. To quantify this effect, we run simulations looking for $N_\text{mode,\,search}$ distortion modes, while we included $N_\text{modes} \geq N_\text{modes,\,search}$ modes in the data. We then look at the error on each estimated coefficient. The simulations are run with noiseless data, 10 frames and a high number of stars ($N_{star}= 1000$) to average the aliasing effect in the whole field.\\
In a first approach, 3rd and lower polynomial orders are estimated ($N_\text{modes, \,search}=10$) while one additional 4th order distortion mode ($b_{11}$, proportional to $\sim x^4$) is introduced on the $x$-axis. This simulation is referred to as \textit{Simulation 5} in Table \ref{tab:simu}.  The errors on each coefficient are shown in Figure. \ref{Figure:spec1}: Green bars represent the errors on the estimated coefficients quadratically averaged on all the images in absolute value, while red bars shows the value of the additional coefficient introduced and non-estimated. For convenience, the modes are identified according to their order in $x$ and $y$ variables. For each mode, the first value corresponds to the $x$-axis distortion and the second value corresponds to the $y$-axis distortion.

We observe here three important results about the aliasing of mode $b_{11}$, proportional to $\sim x^4$ and introduced on the $x$-axis: on average, affected coefficients are all coefficients associated to modes introduced on the $x$-axis. More precisely, affected coefficients are associated to modes proportional to $x^0$ and $x^2$ (i.e., to modes of the same parity as the aliased mode). Finally, the root mean square of the errors match with the value of the additional mode's coefficient: the energy is conserved. 
%The same results are generated for all distortion modes, $m \in [11:21]$.
Those results can be generalized to all distortion modes: the energy is always aliased on modes of similar geometry and the total amount of energy aliased is preserved, which means that there is no amplification of the aliasing noise.
Such results can be used to optimize the algorithm performance: depending on the noise level and on an estimate of the total amount of distortions in the data, we can predict the aliasing error term. By adjusting the number of distortion modes searched, we are able to keep it below a given threshold.
Finally, these results provide precious informations on the effect of different distortions orders, that can be used to better constrain future instrumental designs. 

%Conclure sur l'aliasing : propagation de 1; et suivant le niveau de bruit dans les donnees, on adapte le bombre de mode cherche. Si pas beaucoup de bruit, on peut se permettre de chercher beaucoup de mode. Si beaucoup de bruti, il vaut mieux en chercher moins pour eviter la propagation du bruit dans l'aliasing. Simu pour un bruti donnee, en faisant evoluer le nombre de modes cherche ? 

%[Figure \ref{Figure:spec2} shows the flat spectrum of the $4^{th}$ and $5^{th}$ order polynomials on the x-axis folded on the $3^{th}$ orders polynomials. The additional and non-estimated mode are shown in blue while the errors on the estimated modes are shown in orange. ]---> A enlever ? 

\begin{table*}
\caption{ Summary table of detailed settings used in $Simulations$ 1 to 5. $N_{im}$ is the number of frame, $N_{star}$  is the number of reference sources, $N_\text{modes}$  is the number of distortion modes introduced in the simulated data, $ N_\text{modes,\,search}$ is the number of distortion mode searched in the minimization, $\sigma_{meas}$ is the standard deviation of the Gaussian noise introduced in the position measurements and $n$ the number of outcomes computed for each level of noise.}         
\label{table:simu} 
\begin{tabular}{ccccccc}
   Simulation index & $N_{im}$ & $N_{star}$ & $N_\text{modes}$ &$ N_\text{modes,\,search}$&$\sigma_{meas}$ (px)&$n$ \\
   \hline
   1 & 10 & 30 & 10 &10 & 0.14 &1 \\
         \hline
   2 & 10 & 30 & 10 &10 & [0,0.03,0.07,0.14,0.28] &100\\
      \hline
   3 & [2,10,50,100] & 30 & 10 &10 & [0,0.03,0.07,0.14,0.28]&100\\
      \hline
   4 & 10 & [20,30,100,200] & 10 &10 & [0,0.03,0.07,0.14,0.28] &100\\
   \hline
     5 & 10 & 1000 & 11 &10 & 0.0 &1\\
   \hline
\end{tabular}
\end{table*}

\section{Application to GeMS data}
\label{sec:gems}
In this section, we show the first application of the proposed method on on-sky data. 
The data used are collected with the Gemini MCAO instrument, GeMS combined with the Infra-Red (IR) camera GSAOI (for Gemini South Adaptive Optics Imager). 
%By using multiple Laser Guide Stars (LGS), Wide Field Adaptive Optics (WFAO), improves the performance of high spatial resolution imaging: The AO-corrected images Field of View (FoV) is increased, as well as the fraction of the sky that can benefit from such correction.
The Gemini MCAO instrument GeMS is the first multi-Laser Guide Star (LGS) operational system on sky. It has been implemented on the Gemini South telescope and commissioned in 2013. It operates using two deformable mirrors conjugated at 0 and 9 km and a sodium-based LGS constellation composed of five spots: four are located at the corners of a 60\,arcsec square, with the fifth positioned in the center. GeMS, as a facility instrument, can direct its light output to different science instruments installed at the Cassegrain focus of the Gemini South telescope. Combined with the IR camera GSAOI, it delivers near-diffraction limited images at Near-Infrared (NIR) wavelengths (from 0.9 to 2.4\,$\mu$m) over a FoV of 85$\arcs\times85\arcs$. More details about the GeMS/GSAOI systems and its commissioning results are described in details in previous papers (see \cite{McGregor2004}, \cite{Carrasco2012}, \cite{DOrgeville2012}, \cite{Neichel2013a} and \cite{Rigaut2014}).

In the past few years, the GeMS/GSAOI systems has shown its ability to reach both good astrometric and good photometric precisions. Among the previous studies, we mention here two noteworthy results: \cite{Neichel2014a} show that for single-epoch, un-dithered data, an astrometric error below 0.2\,mas can be achieved for exposure times exceeding one minute, 
and \citet{Turri2015}
%present the deepest \ks \,photometry in crowded fields ever obtained from the ground: a magnitude error of 0.15 at \ks=22 using a total exposure of 1920\,s.
 reached the deepest \ks ~photometry in a crowded field ever obtained from
the ground with a magnitude error of 0.15 at \ks=22 for a total exposure of 1920\,s.

However, regarding to multi-epoch or dithered data, the performance of the GeMS/GSAOI system has been confronted to difficulties due to the presence of static and variable distortions. The origins of these distortions are not totally understood but probable causes have been identified.
%This explain why only few astrometric studies have been using GeMS data (\cite{Turri2015}, \cite{Massari2015}).
First, the off-axis parabola present in the AO bench as well as the deformable mirror conjugated in altitude may introduce low spatial frequency orders of distortions (first and second orders). Then, from the analysis carried in \cite{Neichel2014b}, we know that high spatial frequency are also present (around 15 degrees of freedom). 
Part of these high order distortions may be introduced in dithered data by the gaps between the four GSAOI's detectors. Finally, observational factors such as the Natural Guide Stars (NGS) constellation and the telescope pointing that induces gravity flexure and movement of the AO-bench due to the Cassegrain configuration of the telescope might introduce variable terms of distortions.

GeMS/GSAOI distortions have recently been calibrated by \cite{Massari2016} who used HST data as an external distortion-free reference. The target field is the globular cluster \textit{NGC6681}, which was observed with both systems: WFC/ACS of HST  in 2006, as part of the GO-10775 program (PI : Sarajedini), and GeMS/GSAOI in 2013 as part of the programs GS-2012B-SV-406, GS-2013A-Q-16 and GS-2013B-Q-55 (PI : McConnachie).
They derived from the HST observations the positions of 7770 stars that were then moved according to their Proper Motion (e.g.  \cite{Massari2013}) to construct a distortion-free reference frame at the epoch of the GeMS dataset, 6.914\,years later.

\begin{figure}
\includegraphics[width=\hsize]{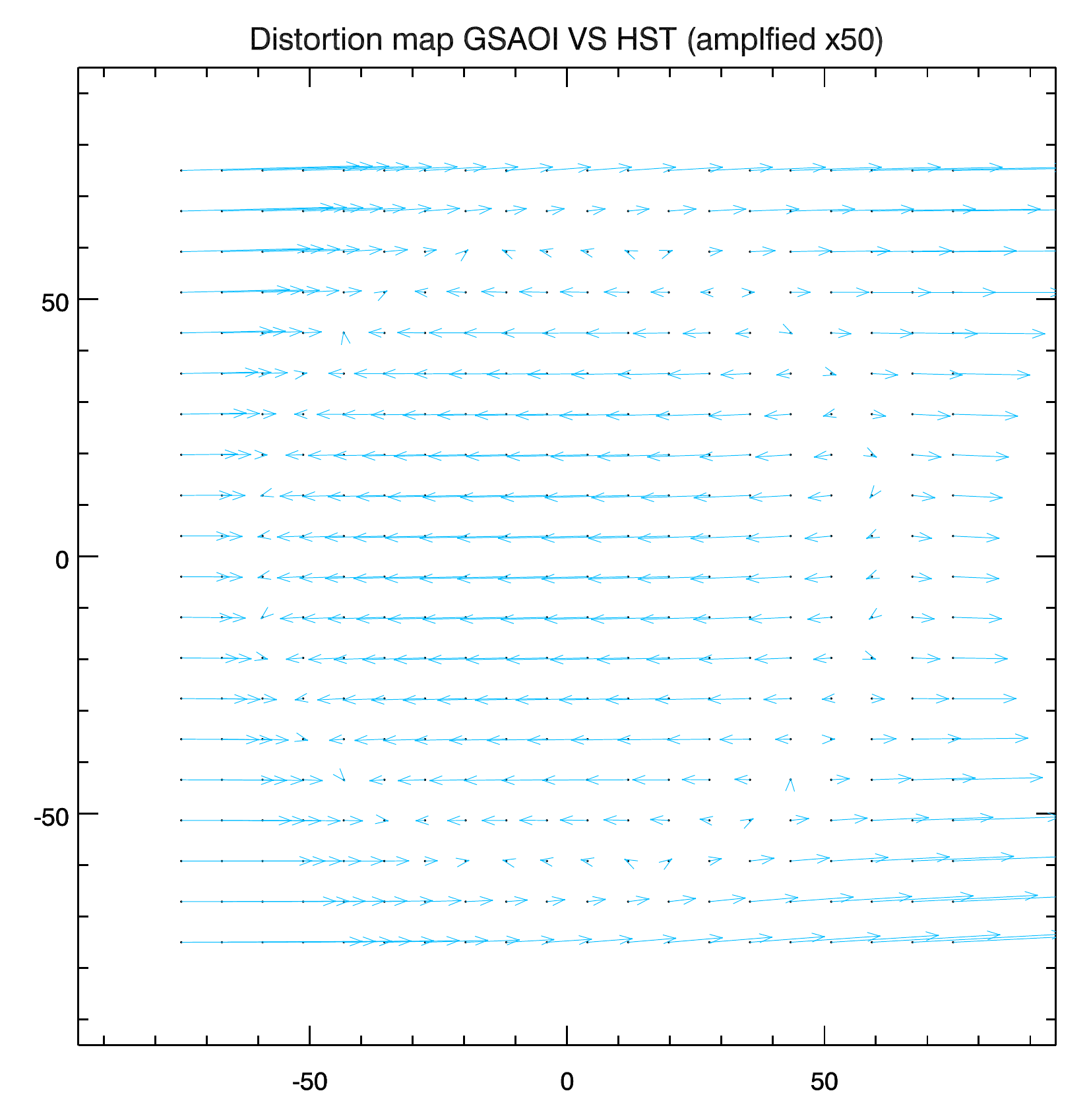}
\caption{GeMS/GSAOI distortion map. The blue arrow show the displacement induced by second and higher distortion orders on one detector. The displacements are amplified by 50.
}
\label{Figure:disto}
\end{figure}

Their processed distortion map, including 
%first-order distortion, shows a peak value in the order of 6\,pixels (120\,mas) on both axis. 
second and higher distortion orders, shows
a peak values of 5\,pixels (100\,mas), on the $x$-component while the $y$-component range below 0.5\,pixels (10\,mas). In addition, circular structures are seen at the center of each chip, where the $x$-component is minimal and then, changes its direction as shown in Figure \ref{Figure:disto}. This distortion solution is derived for the specific data set of the \textit{NGC6681} cluster and can only be applied to this observation set due to the variability of the distortions. However, it represents the best estimation of the GeMS/GSAOI distortion currently available. It is thus used in the following, as input parameter in the method and mentioned as \textit{static distortion map.}

Based on this static distortion map associated to the GeMS/GSAOI instrument, we perform in this Section the first application on real data of the new distortion correction method proposed in this paper. This Section is organized as follows: in a first part, we briefly describe the observations set and data reduction. In a second part, we detail the construction of the star position catalogs.  The third part describes the distortion correction settings and the last part shows the results and highlights the gain brought by this method.

\subsection{Observation and data reduction}
The distortion correction is performed on a GeMS/GSAOI data set extracted from a recent observation of a very active and young star-forming region named \textit{N159W} and located in the Large Magellanic Cloud (LMC). This region was previously observed by  \cite{Bernard2016}, \cite{Deharveng1992}, \cite{Testor2007a}, \cite{Chen2010} in order to study the properties of the cluster stellar members and bring new elements to our understanding of the massive star formation process.
In this study, we use the 11 frames obtained in \ks filter. The \textit{N159W} field provides a large number of well-isolated stars and is therefore an good case to experiment new methods of distortion correction. The data were obtained during the night of December 8th 2014 as part of program GS-2014B-C-2 (P.I. B. Neichel). Each observation consists of one science field dithered randomly by a 5\arcs rms shift to remove gaps between the detectors. The averaged FWHM in \ks band is 90\,mas, while the averaged Strehl ratio (SR) is 14\%. The coordinates of the center of the field are RA = 05h39m40s ; DEC = -69\degree45'55''. \\
The raw images were processed by subtracting dark frames and removing bad pixels from the analysis. Flat-field calibration was performed using twilight sky flats. Sky subtraction was accomplished by forming the median of the dithered frames taken outside of the cluster and subtracting this median from each exposure. As we previously mentioned, the GSAOI camera is composed of 4 chips separated by gaps. In the following, we aim to correct distortion of data collected with the lower left detector as we only have available static distortion for this one chip. Hereafter, it is referred to as detector 2.

\subsection{Construction of input catalogs}

To perform the distortion correction, a list of stellar positions and brightnesses in each exposure is needed, as well as the measurement error $\sigma_{meas,\,j}$ associated to each star $j$. The measurement error will be used for both, the estimation (as it is related to the weighted coefficient $w$ as described in Section \ref{sec:distocorr}) and the quantification of the algorithm performance in Section \ref{sec:results}.
% error will then be used as weighet coefficients and to quantify the permfornance of the algorithm in Section \ref{sec:results}
%This section aims to describe the construction of the input catalogs used in the algorithm. 
The first part of this section is dedicated to the position and flux measurements process. The second part describe the simulations used to estimate the measurement errors.

\subsubsection{Position measurements}
\label{sec:posmeas}
 The stellar positions and brightness catalog is generated using a combination of two star detection tools. Sextractor (\cite{Bertin1996}) is first used to perform a fast detection of the objects in the field and derive position measurements. However, as Sextractor has not been designed to perform accurate astrometry, we use in a second phase a home-made fitting method developed with the interpreted-language Yorick (\cite{Munro1995}), to measure accurate stellar positions. We fit the star intensity distribution using a Moffat profile defined as follows:
\begin{equation}
I = I_0 * \left[1+(X/dx)^2 + (Y/dy)^2\right]^{-\beta} + I_{bkg}
\end{equation}
\noindent where $X = (x-x_0)\cos\theta + (y- y_0)\sin\theta$ and $Y = (y-y_0)\cos\theta - (x-x_0)\sin\theta$. The free parameters of the fit are the positions ($x_0$ and $y_0$), the intensity, ($I_0$), the width in both directions ($dx$ and $dy$), the position angle ($\theta$), and the beta index ($\beta$). The background ($I_{bkg}$) is fitted simultaneously.\\

\begin{figure}
\includegraphics[width=\hsize]{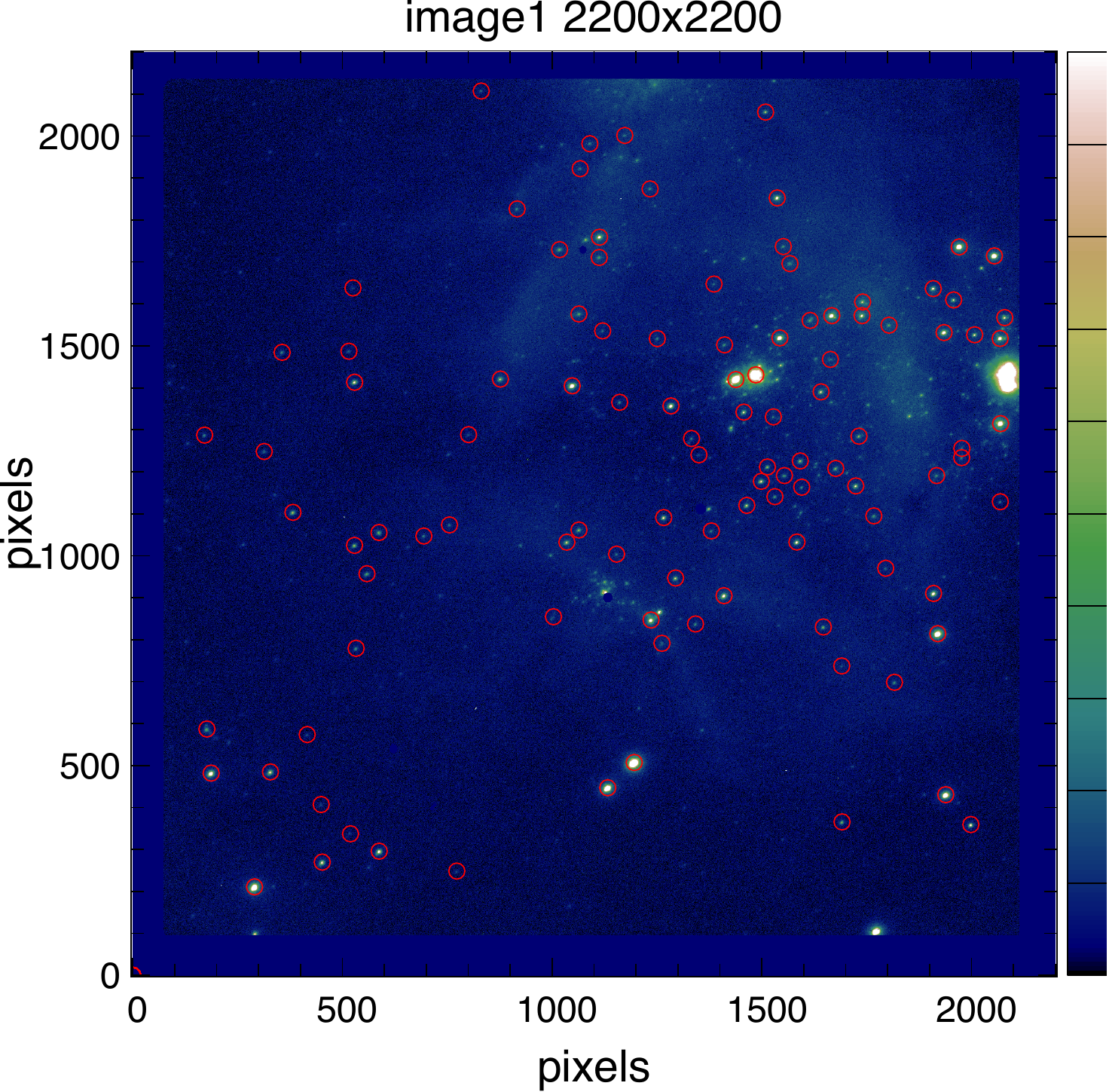}
\caption{One typical reduced frame imaged with the detector 2. Stars used in the minimization are red circled.}
\label{Figure:stars}
\end{figure}

As we need to identify stars positions in each frame, we only consider relatively well-isolated stars in order to avoid any confusion. In addition, any star that did not appear in at least 4 exposures is discarded. This gives us a stellar catalog with positions and flux measurements for each frame. The final number of stars in each catalog range between and 61 and 78. Figure \ref{Figure:stars} shows the lower left part of one reduced frame, corresponding to the part imaged with the detector 2. Stars referenced in the catalog are marked by a red circle.

%Considering that the noise in position measurements is governed by photon noise, the position error is given by the following formula: 
%\begin{equation}
%\sigma \sim \frac{1}{SNR} \sim \frac{1}{\sqrt{Flux}}
%\end{equation}

%The fundamental limit to astrometry for diffraction limited data is given by the number of photons recorded and the image resolution, with higher SNR and smaller point spread function (PSF, FWHM $\Theta$) increasing the positional accu- racy. For a circular aperture of size D and an observation wavelength $\lambda$ one has
%The statistical measurement accuracy is given by
%\begin{equation}
%\Theta_{FWHM} = 1.028 \frac{\lambda}{D}
%\end{equation}
%Lindegren (1978) gives the following formula for the position error:
%\begin{equation}
%\sigma_x=\frac{1}{\pi}\frac{\lambda}{D}\frac{1}{SNR}
%\end{equation}

\subsubsection{Estimation of noise measurements}
\begin{figure}
\includegraphics[width=\hsize]{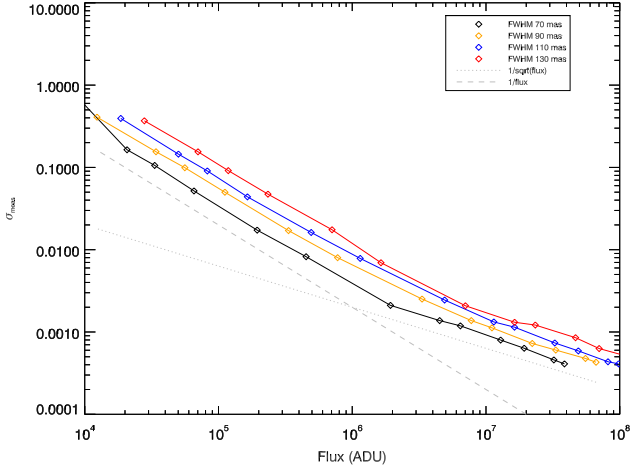}
\caption{Astrometric error (in pixels) versus the flux (in ADU) in specific \textit{N159W} data for different Full Width Half Maximums: 70\,mas (in black), 90\,mas (in orange), 110\,mas (in blue) and 130\,mas in red. The grey dashed line shows the $1/\text{Flux}$ slope and the grey dotted line shows the $1/\sqrt{\text{Flux}}$  slope.}
\label{Figure:noiseflux}
\end{figure}

To estimate the noise measurement, in the specific case of the \textit{N159W} observations, we use simulations. 
The idea is to simulate and ideal Moffat PSF and to embed it, at a known position, in a background image derived from the \textit{N159W} \ks-band data. Then the position is estimated again using the position measurement process described in the previous section (\ref{sec:posmeas}). The measurement error appears immediately as the distance between the real position of the PSF and the fitted position of the PSF.   
The detector noise is already included in the background and the photon noise is added using a Poisson distribution. Different PSF flux levels are explored by scaling the PSF before the photon-noise computation and for each PSF flux level, we simulated a set of 100 images with different known positions of the PSF. %Finally, we determined this position using a Moffat fit to each image and compute positional uncertainty. 
The positional uncertainties are then quadratically averaged on the 100 positions for each intensity level.\\
Results are presented in Figure \ref{Figure:noiseflux}. The solid lines shows the astrometric error (in pixels) versus the flux (in Analog-to-Digital Units (ADU)). The four curves corresponds to four different Full Width Half Maximum (FWHM) of the simulated PSF: 70, 90, 110 and 130\,mas ( respectively in black, orange, blue and red line). The grey dashed line shows the $1/\text{Flux}$ slope and the grey dotted line shows the $1/\sqrt{\text{Flux}}$ slope. These two slopes highlights two regimes: a $1/\text{Flux}$ evolution for fluxes lower than $10^6$ ADU and a $1/\sqrt{\text{Flux}}$ for higher fluxes. The former regime is dominated by the detector and sky noise, the latter being dominated by the PSF photon noise. 
%These results are very consistent with those derived by \citet{Fritz2010}.

\subsection{Distortion correction}
\label{sec:distocorr}
The distortion correction is then performed using the following settings (summarized in Table \ref{tab:simu}):
\begin{table*}
\caption{ Summary table of settings used in distortion correction of \textit{N159W} data. $N_{im}$ is the number of frame, $N_{star}$  is the number of reference sources, $ N_\text{modes,\,search}$ is the number of distortion mode searched in the minimization, $w$ is the weight coefficient and $ [A\,]^0$ the initialization of the distortion coefficients set according to the prior knowledge of the static distortion}         
\label{tab:simu}
\begin{tabular}[width=\hsize]{ccccccc}
    $N_{im}$ & $N_{star}$ & $ N_\text{modes,\,search}$&d&$w$ &$ [A\,]^0$ \\
   \hline
   11 &$  [61:78] $& 10 & 3&$1/\sigma_{meas}^2$ &  \cite{Massari2016} static distortion solution\\
   &&&&& fitted on 2D Legendre normed basis\\
   \hline
\end{tabular}
\end{table*} 
The weighting coefficient $w_{i,\,j}$ is set for each star $j$ according to its positional uncertainty $\sigma_{meas,j}$. The positional uncertainty is derived from the flux of each star according to the simulation curve presented in the last section for FWHM=90\,mas, which is the averaged FWHM in the data. It ranges between 0.002\,pixels and 0.14\,pixels (respectively 0.02\,mas and 2.8\,mas), mostly depending on the magnitude of the objects. 
\begin{equation}
w_{i,\,j}=\frac{1}{\sigma_{meas,j}^2}
\end{equation}

We recall here that the distorsion map derived for the specific data set of \textit{NGC6681} by \cite{Massari2016}  is the best estimation of the static distorsion present in the GeMS/GSAOI instrument currently available.
The initial distortion coefficients $[A\,]^0$ are then derived from this static distortion solution fitted on the 2D Legendre polynomials normed basis.
As the fit does not improve beyond 10 distortion modes considered on both $x$ and $y$ axis, we use this setting to perform the distortion correction: $N_\text{mode,search}= 10$ which corresponds to 3rd order polynomials ($d=3$).

\subsubsection{Results}
\label{sec:results}
\begin{figure}
\includegraphics[width=\hsize]{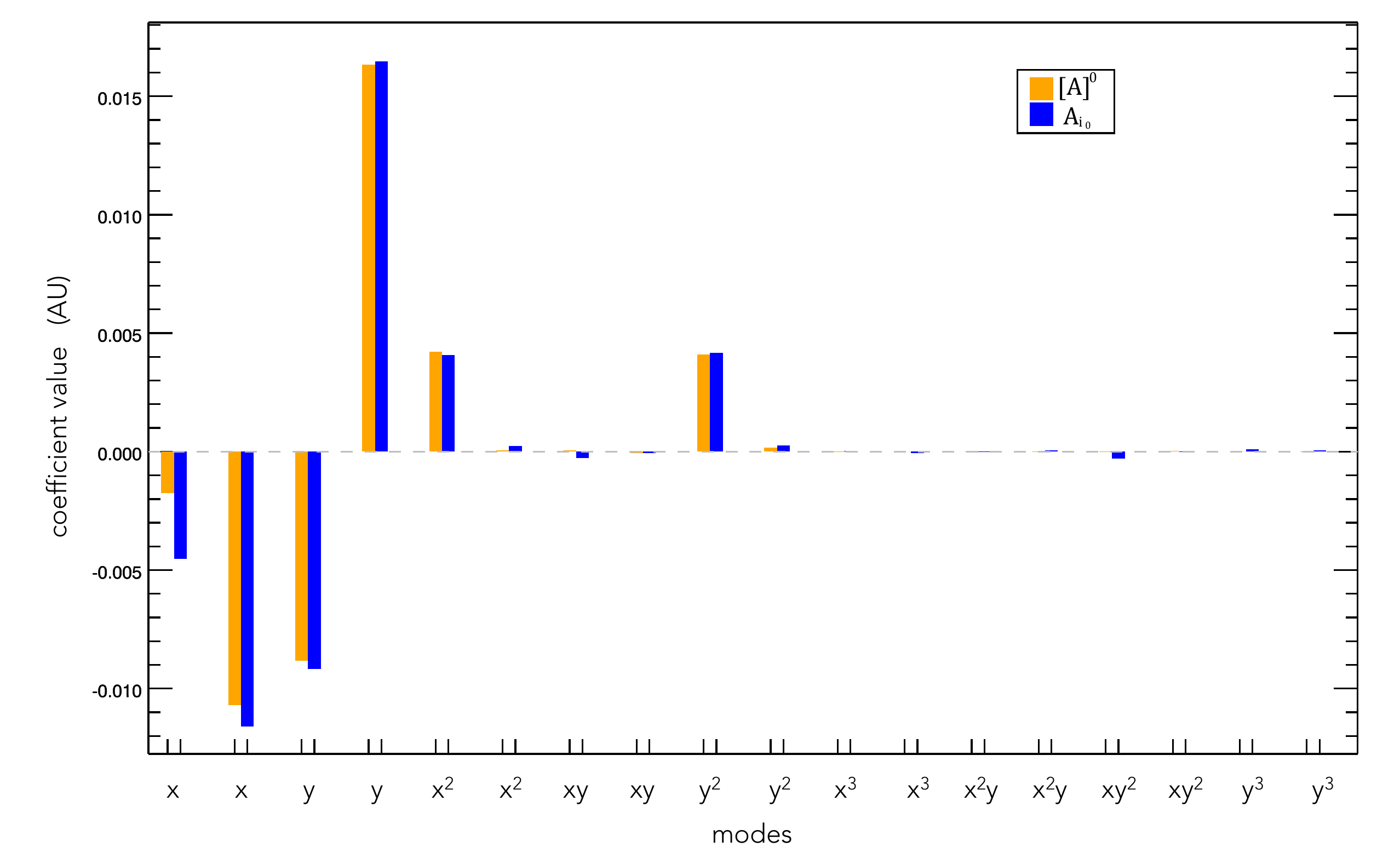}
\caption{Bargraph of the estimated distortion coefficient $A_{i_0}$ associated to frame $i_0$ (in blue bars) and the distortion coefficients associated to the static distortion solution, $[A\,]^0$ (in orange bars). The modes are identified according to their order in $x$ and $y$ variables. For each mode, the first value corresponds to the $x$-axis distortion and the second value corresponds to the $y$-axis distortion.
}
\label{Figure:coeff}
\end{figure}

The distortion solution derived for each frame is varying around the distortion solution provided by \cite{Massari2016}. For exemple, Figure \ref{Figure:coeff} shows the estimated distortion coefficient value $A_{i_0}$ of frame $i_0$ (in blue bars) and the distortion coefficients derived from the static distortion solution, $[A\,]^0$ (in orange bars). 
%On this Figure, the modes are identified according to their order in $x$ and $y$ variables. For each mode, the first value corresponds to the $x$-axis distortion and the second value corresponds to the $y$-axis distortion.

\begin{figure}
\includegraphics[width=\columnwidth]{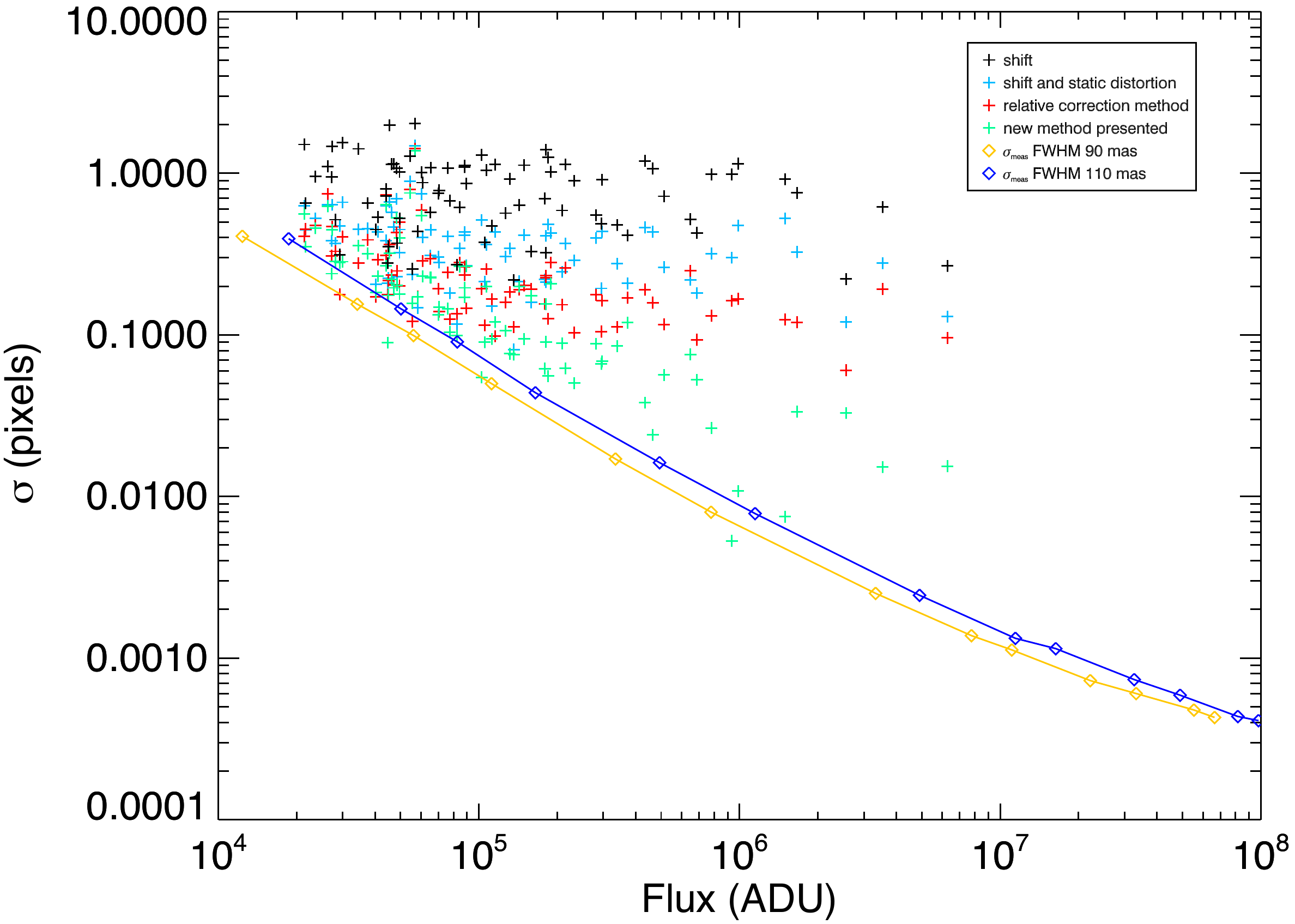}
\caption{Corrected position standard deviation of each star ($\sigma_j$ in pixel) as function of its flux (ADU) for different type of corrections: A shift correction (in black). A static distortion correction using the static distortion map and a shift correction (in blue), a relative distorsion correction (red), and a distortion correction performed with the method preposed (in green). Orange and blue curves represent the astrometric error versus the flux calibrated in GeMS/GSAOI data corresponding to a 90\,mas and 110\,mas FWHM respectively. %Grey dashed line shows the $1/Flux$ slope and grey dotted line shows the $1/\sqrt{Flux}$ slope}
}\label{Figure:ercomp}
\end{figure}

In order to estimate the correction efficiency we plot the corrected position standard deviation of each star ($\sigma_j$) as function of its flux. %It corresponds to the variance part of the $rms_{astro}$ defined previously.
%To compute this error, one way to do would be to re detect all the stars in the corrected frame using a detection software and compute the error using the new positions detected. One other way is to compute the correction is directly to the data positions. In this way the results are immediate and a detection term error is not add to the budget.
Each star is represented by a green cross on Figure. \ref{Figure:ercomp}. 
As a comparison, we also plot the same quantities obtained by performing a simple shift correction (black), by performing a correction of the static distortion in addition to the shift correction (blue), and by performing a relative distorsion correction as it was originally done in \cite{Bernard2016} using the same data set (red).
The static distortion is removed by applying to each frame the static distortion map derived by \cite{Massari2016} and the relative distorsion correction is applied following the process described in detail in \cite{Bernard2016}.
On Figure \ref{Figure:ercomp}, the curves shows the position uncertainty as function of the flux for a 90\,mas FWHM star (orange line) and a 110\,mas FWHM star (blue line). Those are the curves calculated in Section \ref{sec:posmeas} specifically for the \textit{N159W} data. 
%The grey dashed line shows the $1/Flux$ slope and the grey dotted line shows the $1/\sqrt{Flux}$ slope.

The Figure \ref{Figure:ercomp} illustrates the gain brought by the different methods : the static distortion correction improves the astrometric precision to an averaged $\sigma_\text{static disto.}=0.41$\,pixels and a standard deviation of $0.21$\,pixels, compare to a simple shift correction where $\sigma_\text{shift}=0.83$\,pixels and the standard deviation is 0.4\,pixels. However, the static distortion map derived for the specific data set of the \textit{NGC6681} cluster is not perfectly suited for any other set of observation due to the variability of the distortions. 
By applying a relative correction of distortion, these variabilities are taken into account as each frame is corrected individually according to one reference 
frame chosen arbitrary within the data. The astrometric error then falls down to an averaged precision of $\sigma_\text{new meth.}=0.24$\,pixels, and a standard deviation of $0.21$\,pixels.

The method proposed in this paper presents both advantages to handles the distorsion variabilities by adapting distortion maps to each set of observations and each individual frame, and to provide an optimal noise propagation thanks to the weighted coefficient. In this way, the error term due to distortion is minimized and the astrometric performance is further improved to an averaged precision of $\sigma_\text{new meth.}=0.20$\,pixels, and a standard deviation of $0.20$\,pixels. Plus, we recall here that an additional advantage of the proposed method is to provide an estimation of the distortion-free reference.
In the case of a perfect optical system, the astrometric error is determined by the measurement noise.
This means that any source of additional error needs to be compensated down to this level if we want to fully exploit the data capabilities.
That is what we achieve by performing the method presented in this paper: the distortion error term is minimized and the resulting astrometric uncertainties are dominated by noise measurement. Of course, the final astrometric performance still depends on a large panel of parameters, such as the seeing conditions, the FWHM, the nature of the field : crowded or sparse, the presence of background structures, etc. However, the proposed method guaranties an optimal minimization of the distortion error term. Plus, the \textit{weighted} property of the minimization provides an ideal noise management:  this allows the use of all reference sources available, even the faintest ones, which is an advantage for sparse field studies.

%The gain bought by the \textit{correction 2} over \textit{correction 1} is du to the dithering. As the correction of the static distortion consists in applying the same operation on all the frames, it does not influe the standard deviation of the stars positions through the stack on undithered data.

\begin{figure}
\includegraphics[width=\hsize]{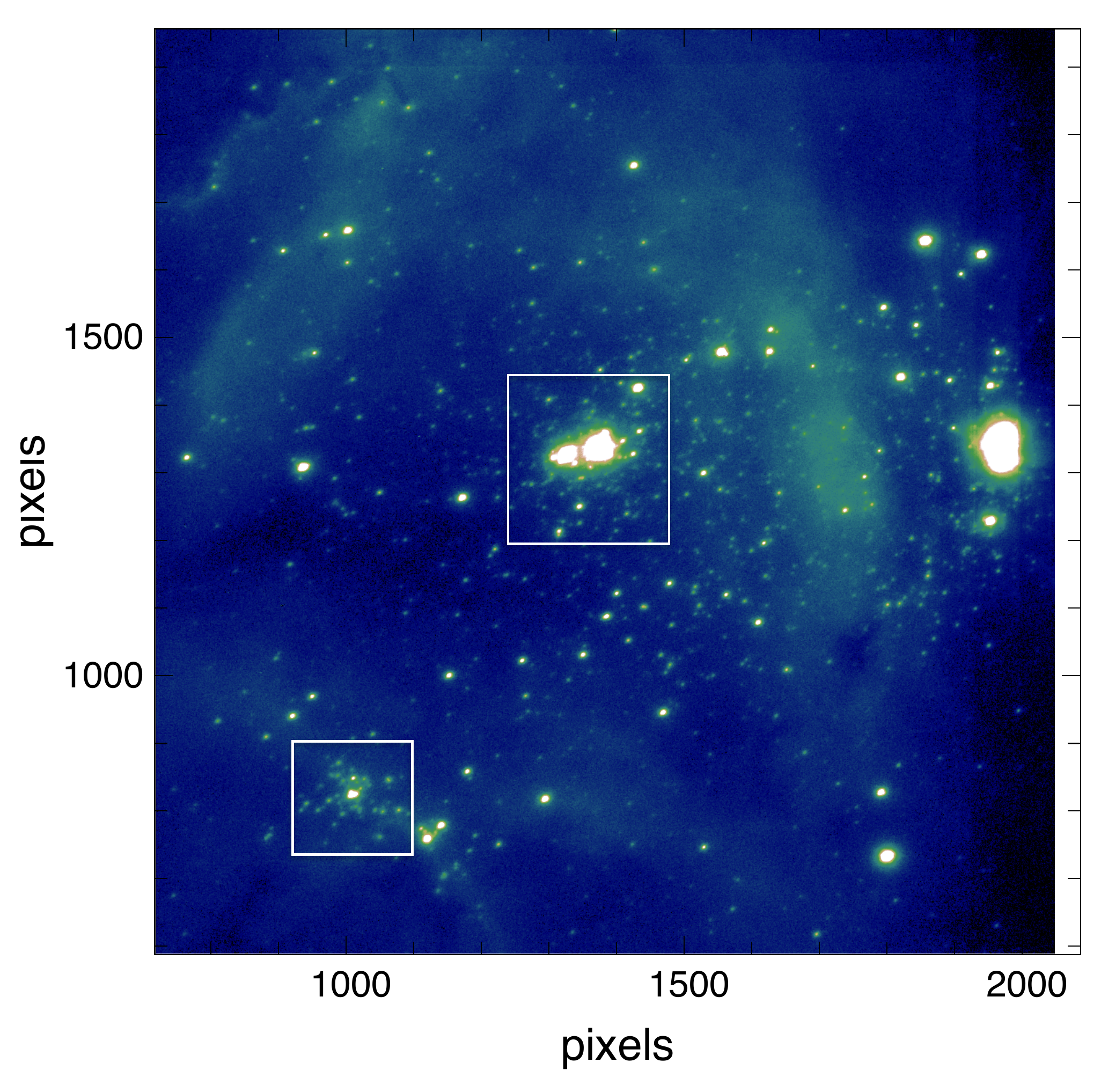}
\caption{Final stack on the 11 frames individually corrected from distortion by performing the optimal distortion correction presented. White squares show parts of the fields that are zoomed in and shown in Figure \ref{Figure:stars2}.
}
\label{Figure:im}
\end{figure}

\begin{figure}
\includegraphics[width=0.5\hsize]{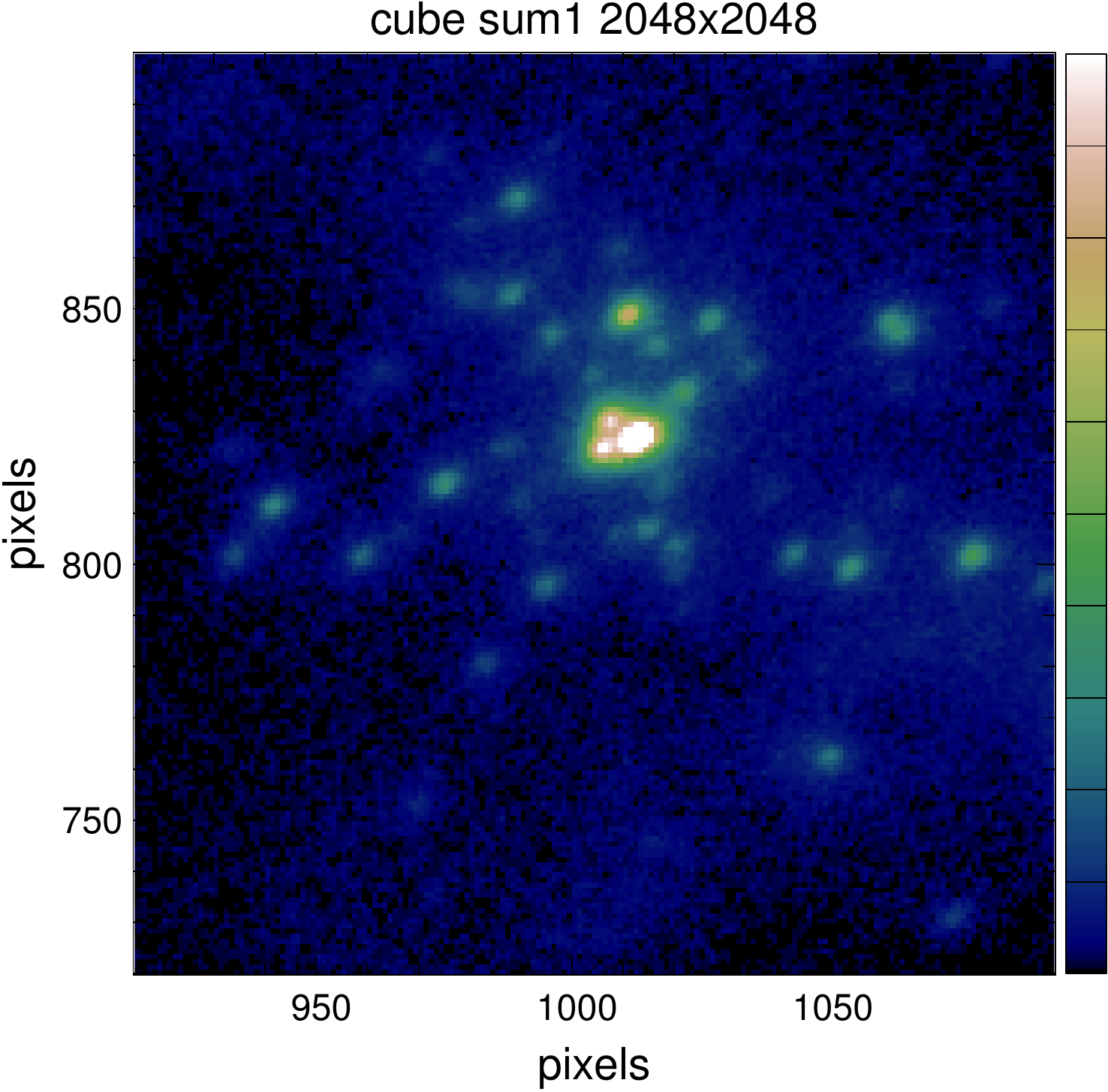}
\includegraphics[width=0.5\hsize]{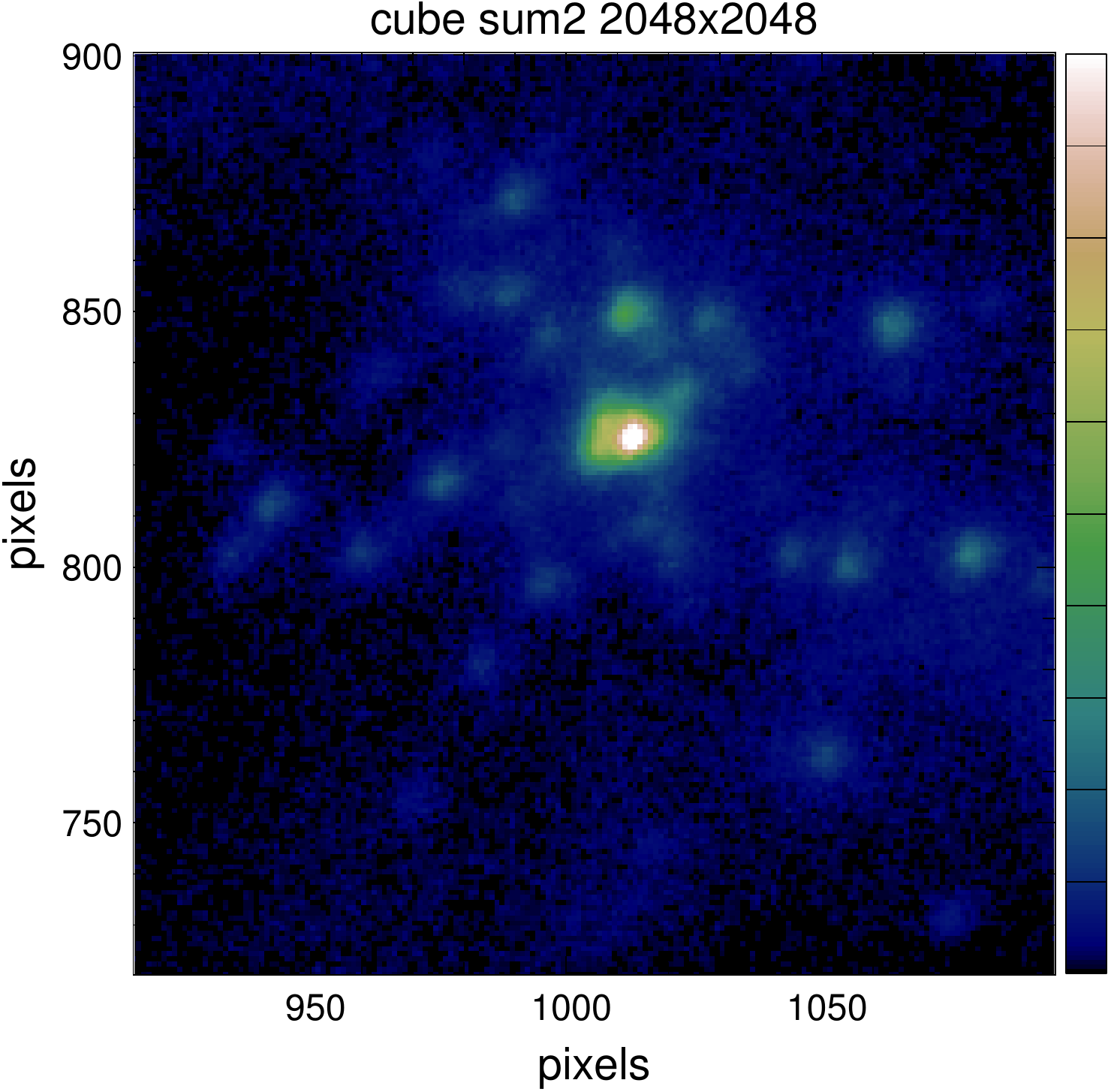}
\includegraphics[width=0.5\hsize]{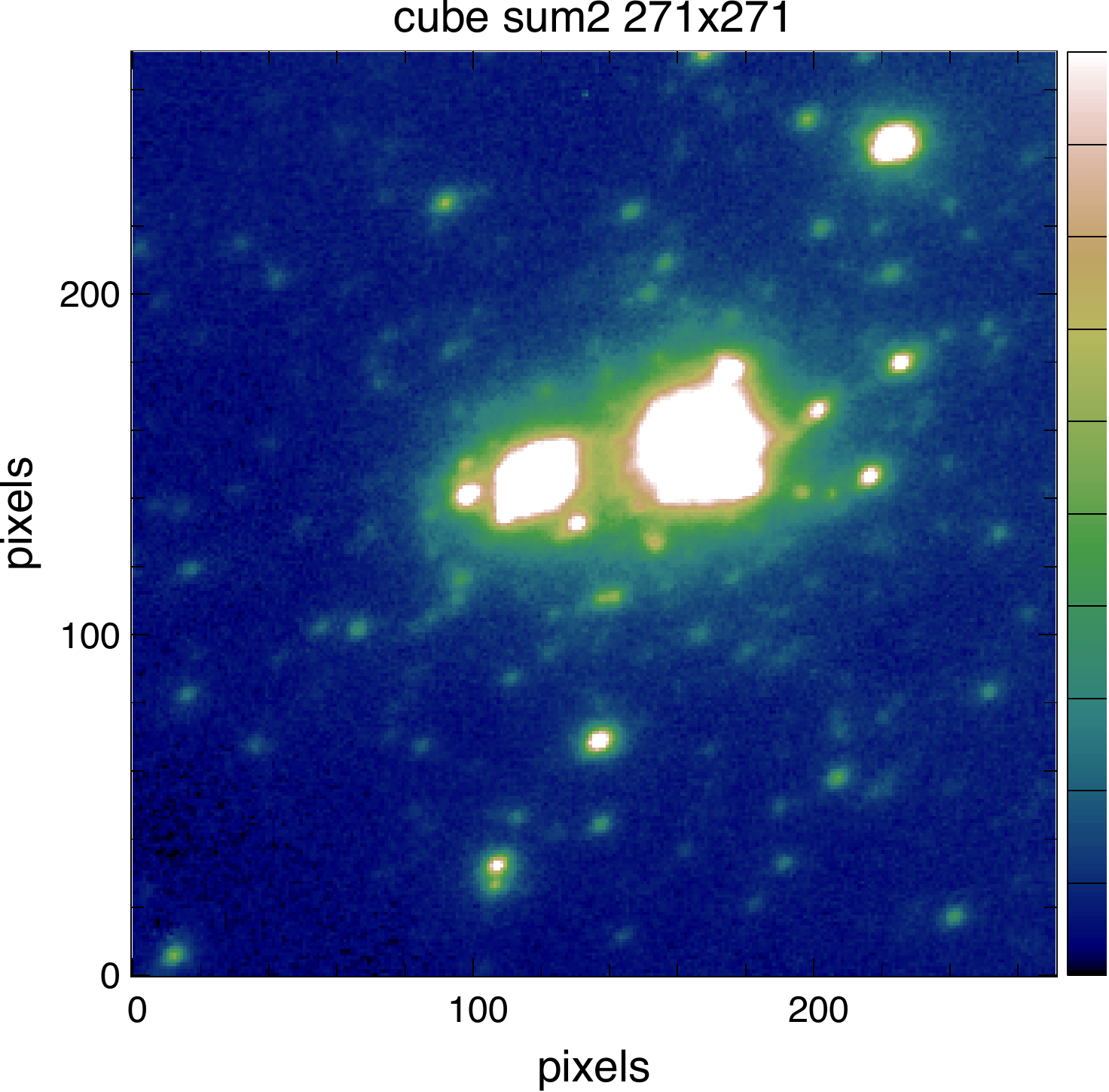}
\includegraphics[width=0.5\hsize]{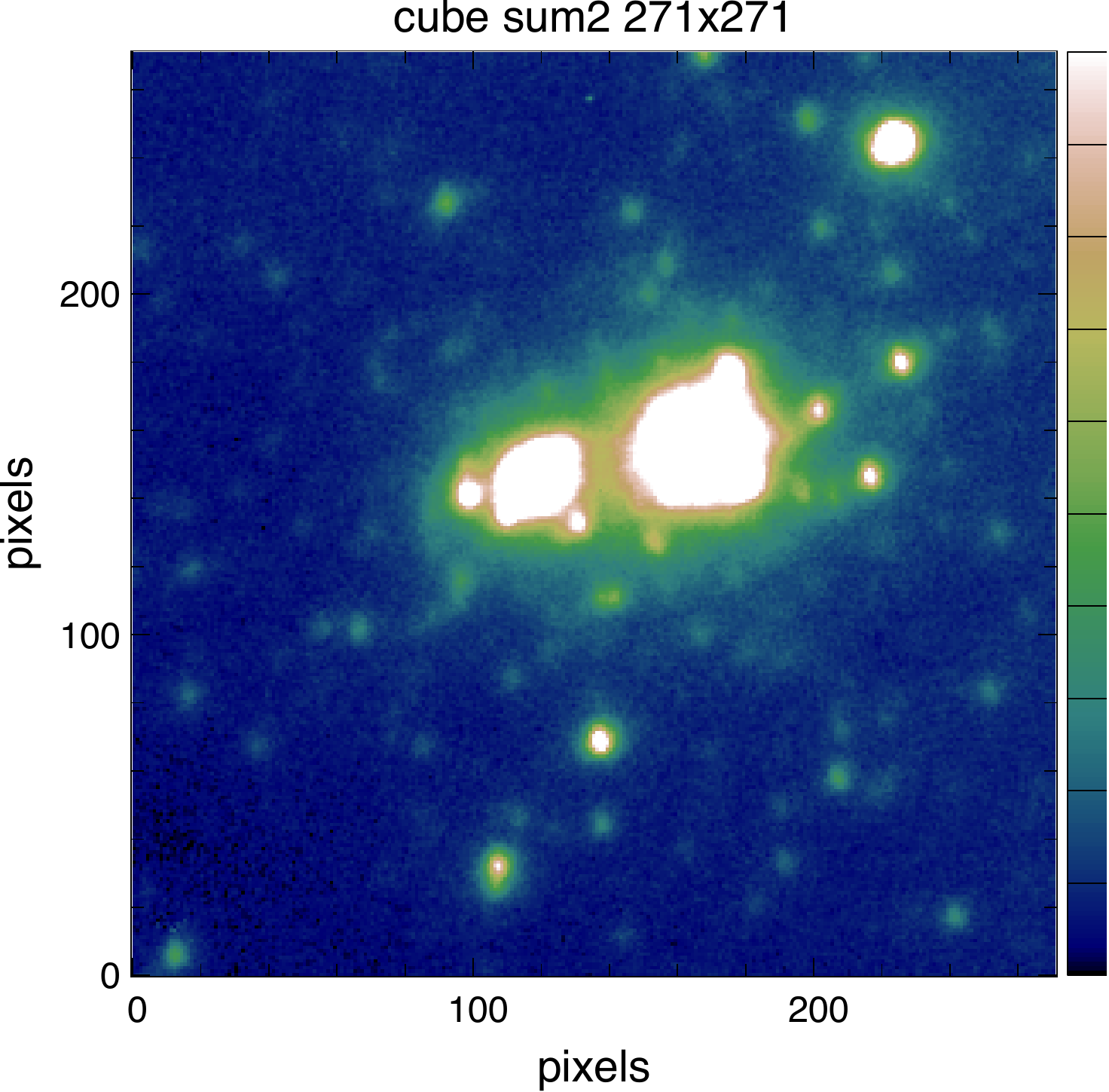}
\caption{Part of the final stack image obtained by performing the the optimal distortion correction presented (left panel) and a static distortion correction (right panel).
 Note that the brightest regions are saturated in order to better show the resolution improvement brought by our method on the faintest stars.}
\label{Figure:stars2}
\end{figure}

As a final step, the correction is applied to each of the 11 frames by an interpolation process using the estimated distortion coefficient $[\widehat{A}\,]$. The final stacked image is shown in Figure \ref{Figure:im}. Parts of the field (indicated by a white square) are zoomed in and shown in Figure \ref{Figure:stars2}. This Figure shows results obtained by performing the optimal distortion correction presented (left panel) and a static distortion correction (right panel). Note that the brightest regions are saturated in order to better show the resolution improvement brought by our method on the faintest stars.

%
%ta méthode est optimale. Donc peu importe les données, et peut importe la qualité des données, c'est ta méthode qui permettra la meilleure estimation des paramètres. Parcequ'elle est optimale. D'ailleurs, tu pourrais plus insister sur les aspects "on peut aller chercher meme les étoiles très faibles, donc pour les champs sparse, c'est parfaitement adapté", ou sur des choses du genre "ont a une gestion optimale du bruit, avec le coefficient "wi" donc on ne propagera jamais du mauvais bruit, donc on gagnera toujours à utiliser un maximum d'étoiles". 

\section{Conclusion}

In this paper we presented an new and optimal method of distortion correction for high angular resolution images. Based on a prior-knowledge of the static distortion present in the data, this method aims to correct the dynamical distortions by specifically adapting to each set of observation and each frame. The method consists in an inverse problem approach solved by a Weighted Least Square minimisation. Thanks to the \textit{weighted} property of the minimization, we guaranty a minimal noise propagation that  allows the use of the faintest reference sources as algorithm input. This is a very valuable advantage if working on sparse fields images.
The resolution provides an estimation of both a distortion-free reference and a distortion solution associated to each frame. 
Depending on the scientific aim of the study (astrometry, photometry), both parameters can be used independently. 
We presented the complete algorithm implementation as well as simulations studies aiming to characterize the method performance facing different type of observations: crowded or sparse field, varying number of frame, positions of the reference sources, number of distortion modes. Those results allow us to optimize the algorithm settings regarding each configuration of data and scientific goal, and to predict the expected performance for a given observation set.
Finally, we show the first application of the method using on-sky data collected with the Gemini MCAO instrument, GeMS. We reach an astrometric precision of 0.2\,pixels which represents a gain of factor two compared to a classical static distortion correction and a factor 1.2 compare to a relative distorsion correction.
The final astrometric performance still depends on a large number of parameters, such as the seeing conditions, the FWHM, the nature of the field : crowded or sparse or the presence of background structures. However, the proposed method guaranties an optimal minimization of the distortion error term which conducts to a resulting astrometric error falling down to the noise measurement level. 

%\begin{verbatim}
%\documentclass[onecolumn]{mnras}
%\end{verbatim}

%The simulation of astrometric precision afforded by the optimal weighting scheme illustrates that measurement noise is the dominant residual astrometric error on a 8 meter telescope for stellar fields that contain more than a few reference stars.

\section*{Acknowledgements}
Based on observations obtained at the Gemini Observatory, which is operated by the Association of Universities for Research in Astronomy, Inc., under a cooperative agreement with the NSF on behalf of the Gemini partnership: the National Science Foundation (United States), the National Research Council
(Canada), CONICYT (Chile), the Australian Research Council (Australia), Minist\'erio da Ci$\rm{\hat{e}}$ncia, Tecnologia
e Inova{\c{c}}$\rm{\tilde{a}}$o (Brazil) and Ministerio de Ciencia, Tecnologiaa e Innovaci\'on Productiva (Argentine).\\ 
B. Neichel and A. Bernard acknowledge the financial support from the French ANR program WASABI to carry out this work.\\The authors are grateful to Dr Carrasco for careful reading of the paper and valuable suggestions and comments.\\

\bibliographystyle{mnras}

\bibliographystyle{mnras}

\bibliography{library_ab_mnras.bib} % if your bibtex file is called example.bib

\begin{thebibliography}{}
\makeatletter
\relax
\def\mn@urlcharsother{\let\do\@makeother \do\$\do\&\do\#\do\^\do\_\do\%\do\~}
\def\mn@doi{\begingroup\mn@urlcharsother \@ifnextchar [ {\mn@doi@}
  {\mn@doi@[]}}
\def\mn@doi@[#1]#2{\def\@tempa{#1}\ifx\@tempa\@empty \href
  {http://dx.doi.org/#2} {doi:#2}\else \href {http://dx.doi.org/#2} {#1}\fi
  \endgroup}
\def\mn@eprint#1#2{\mn@eprint@#1:#2::\@nil}
\def\mn@eprint@arXiv#1{\href {http://arxiv.org/abs/#1} {{\tt arXiv:#1}}}
\def\mn@eprint@dblp#1{\href {http://dblp.uni-trier.de/rec/bibtex/#1.xml}
  {dblp:#1}}
\def\mn@eprint@#1:#2:#3:#4\@nil{\def\@tempa {#1}\def\@tempb {#2}\def\@tempc
  {#3}\ifx \@tempc \@empty \let \@tempc \@tempb \let \@tempb \@tempa \fi \ifx
  \@tempb \@empty \def\@tempb {arXiv}\fi \@ifundefined
  {mn@eprint@\@tempb}{\@tempb:\@tempc}{\expandafter \expandafter \csname
  mn@eprint@\@tempb\endcsname \expandafter{\@tempc}}}

\bibitem[\protect\citeauthoryear{{Abuter}, {Schreiber}, {Eisenhauer}, {Ott},
  {Horrobin}  \& {Gillesen}}{{Abuter} et~al.}{2006}]{Abuter2006}
{Abuter} R.,  {Schreiber} J.,  {Eisenhauer} F.,  {Ott} T.,  {Horrobin} M.,
  {Gillesen} S.,  2006, \mn@doi [\nar] {10.1016/j.newar.2006.02.008}, \href
  {http://adsabs.harvard.edu/abs/2006NewAR..50..398A} {50, 398}

\bibitem[\protect\citeauthoryear{{Anderson} \& {King}}{{Anderson} \&
  {King}}{2003}]{Anderson2003}
{Anderson} J.,  {King} I.~R.,  2003, \mn@doi [\pasp] {10.1086/345491}, \href
  {http://adsabs.harvard.edu/abs/2003PASP..115..113A} {115, 113}

\bibitem[\protect\citeauthoryear{{Anderson} \& {King}}{{Anderson} \&
  {King}}{2004}]{Anderson2004}
{Anderson} J.,  {King} I.~R.,  2004, Technical report, {Multi-filter PSFs and
  Distortion Corrections for the HRC}

\bibitem[\protect\citeauthoryear{{Anderson}, {Bedin}, {Piotto}, {Yadav}  \&
  {Bellini}}{{Anderson} et~al.}{2006}]{Anderson2006}
{Anderson} J.,  {Bedin} L.~R.,  {Piotto} G.,  {Yadav} R.~S.,   {Bellini} A.,
  2006, \mn@doi [\aap] {10.1051/0004-6361:20065004}, \href
  {http://adsabs.harvard.edu/abs/2006A%26A...454.1029A} {454, 1029}

\bibitem[\protect\citeauthoryear{Bellini \& Bedin}{Bellini \&
  Bedin}{2009}]{Bellini2009}
Bellini a.,  Bedin L.~R.,  2009, \mn@doi [\pasp] {10.1086/649061}, 121, 20

\bibitem[\protect\citeauthoryear{{Bellini}, {Bedin}, {Piotto}, {Milone},
  {Marino}  \& {Villanova}}{{Bellini} et~al.}{2010}]{Bellini2010}
{Bellini} A.,  {Bedin} L.~R.,  {Piotto} G.,  {Milone} A.~P.,  {Marino} A.~F.,
  {Villanova} S.,  2010, \mn@doi [\aj] {10.1088/0004-6256/140/2/631}, \href
  {http://adsabs.harvard.edu/abs/2010AJ....140..631B} {140, 631}

\bibitem[\protect\citeauthoryear{{Bernard}, {Neichel}, {Samal}, {Zavagno},
  {Andersen}, {Evans}, {Plana}  \& {Fusco}}{{Bernard}
  et~al.}{2016}]{Bernard2016}
{Bernard} A.,  {Neichel} B.,  {Samal} M.~R.,  {Zavagno} A.,  {Andersen} M.,
  {Evans} C.~J.,  {Plana} H.,   {Fusco} T.,  2016, \mn@doi [\aap]
  {10.1051/0004-6361/201628754}, \href
  {http://adsabs.harvard.edu/abs/2016A%26A...592A..77B} {592, A77}

\bibitem[\protect\citeauthoryear{{Bertin} \& {Arnouts}}{{Bertin} \&
  {Arnouts}}{1996}]{Bertin1996}
{Bertin} E.,  {Arnouts} S.,  1996, \mn@doi [\aaps] {10.1051/aas:1996164}, \href
  {http://adsabs.harvard.edu/abs/1996A%26AS..117..393B} {117, 393}

\bibitem[\protect\citeauthoryear{{Cameron} \& {Kulkarni}}{{Cameron} \&
  {Kulkarni}}{2007}]{Cameron2007a}
{Cameron} P.~B.,  {Kulkarni} S.~R.,  2007, in American Astronomical Society
  Meeting Abstracts. p.~996

\bibitem[\protect\citeauthoryear{{Cameron}, {Britton}  \& {Kulkarni}}{{Cameron}
  et~al.}{2009}]{Cameron2009}
{Cameron} P.~B.,  {Britton} M.~C.,   {Kulkarni} S.~R.,  2009, \mn@doi [\aj]
  {10.1088/0004-6256/137/1/83}, \href
  {http://adsabs.harvard.edu/abs/2009AJ....137...83C} {137, 83}

\bibitem[\protect\citeauthoryear{{Carrasco} et~al.,}{{Carrasco}
  et~al.}{2012}]{Carrasco2012}
{Carrasco} E.~R.,  et~al., 2012, in Adaptive Optics Systems III. p. 84470N,
  \mn@doi{10.1117/12.926240}

\bibitem[\protect\citeauthoryear{{Chen} et~al.,}{{Chen}
  et~al.}{2010}]{Chen2010}
{Chen} C.-H.~R.,  et~al., 2010, \mn@doi [\apj] {10.1088/0004-637X/721/2/1206},
  \href {http://adsabs.harvard.edu/abs/2010ApJ...721.1206C} {721, 1206}

\bibitem[\protect\citeauthoryear{{Deharveng}, {Caplan}  \&
  {Lombard}}{{Deharveng} et~al.}{1992}]{Deharveng1992}
{Deharveng} L.,  {Caplan} J.,   {Lombard} J.,  1992, \aaps, \href
  {http://adsabs.harvard.edu/abs/1992A%26AS...94..359D} {94, 359}

\bibitem[\protect\citeauthoryear{Dunkl \& Xu}{Dunkl \&
  Xu}{2014}]{dunkl2014orthogonal}
Dunkl C.,  Xu Y.,  2014, Orthogonal Polynomials of Several Variables.
Encyclopedia of Mathematics and its Applications, Cambridge University Press,
  \url {https://books.google.fr/books?id=eRdEBAAAQBAJ}

\bibitem[\protect\citeauthoryear{{Fritz} et~al.,}{{Fritz}
  et~al.}{2010}]{Fritz2010}
{Fritz} T.,  et~al., 2010, \mn@doi [\mnras] {10.1111/j.1365-2966.2009.15707.x},
  \href {http://adsabs.harvard.edu/abs/2010MNRAS.401.1177F} {401, 1177}

\bibitem[\protect\citeauthoryear{{Ghez} et~al.,}{{Ghez}
  et~al.}{2008}]{Ghez2008}
{Ghez} A.~M.,  et~al., 2008, \mn@doi [\apj] {10.1086/592738}, \href
  {http://adsabs.harvard.edu/abs/2008ApJ...689.1044G} {689, 1044}

\bibitem[\protect\citeauthoryear{{Gratadour}, {Mugnier}  \&
  {Rouan}}{{Gratadour} et~al.}{2005}]{Gratadour2005}
{Gratadour} D.,  {Mugnier} L.~M.,   {Rouan} D.,  2005, \mn@doi [\aap]
  {10.1051/0004-6361:20042188}, \href
  {http://adsabs.harvard.edu/abs/2005A%26A...443..357G} {443, 357}

\bibitem[\protect\citeauthoryear{{Hasan} \& {Bely}}{{Hasan} \&
  {Bely}}{1994}]{Hasan1994}
{Hasan} H.,  {Bely} P.~Y.,  1994, in {Hanisch} R.~J.,  {White} R.~L.,  eds, The
  Restoration of HST Images and Spectra - II. p.~157

\bibitem[\protect\citeauthoryear{{Libralato}, {Bellini}, {Bedin}, {Piotto},
  {Platais}, {Kissler-Patig}  \& {Milone}}{{Libralato}
  et~al.}{2014}]{Libralato2014}
{Libralato} M.,  {Bellini} A.,  {Bedin} L.~R.,  {Piotto} G.,  {Platais} I.,
  {Kissler-Patig} M.,   {Milone} A.~P.,  2014, \mn@doi [\aap]
  {10.1051/0004-6361/201322059}, \href
  {http://adsabs.harvard.edu/abs/2014A%26A...563A..80L} {563, A80}

\bibitem[\protect\citeauthoryear{{Lu}, {Ghez}, {Hornstein}, {Morris}, {Becklin}
   \& {Matthews}}{{Lu} et~al.}{2009}]{Lu2009}
{Lu} J.~R.,  {Ghez} A.~M.,  {Hornstein} S.~D.,  {Morris} M.~R.,  {Becklin}
  E.~E.,   {Matthews} K.,  2009, \mn@doi [\apj] {10.1088/0004-637X/690/2/1463},
  \href {http://adsabs.harvard.edu/abs/2009ApJ...690.1463L} {690, 1463}

\bibitem[\protect\citeauthoryear{{Massari}, {Bellini}, {Ferraro}, {van der
  Marel}, {Anderson}, {Dalessandro}  \& {Lanzoni}}{{Massari}
  et~al.}{2013}]{Massari2013}
{Massari} D.,  {Bellini} A.,  {Ferraro} F.~R.,  {van der Marel} R.~P.,
  {Anderson} J.,  {Dalessandro} E.,   {Lanzoni} B.,  2013, \mn@doi [\apj]
  {10.1088/0004-637X/779/1/81}, \href
  {http://adsabs.harvard.edu/abs/2013ApJ...779...81M} {779, 81}

\bibitem[\protect\citeauthoryear{{Massari} et~al.,}{{Massari}
  et~al.}{2016}]{Massari2016}
{Massari} D.,  et~al., 2016, \mn@doi [\aap] {10.1051/0004-6361/201629336},
  \href {http://adsabs.harvard.edu/abs/2016A%26A...595L...2M} {595, L2}

\bibitem[\protect\citeauthoryear{McGregor et~al.,}{McGregor
  et~al.}{2004}]{McGregor2004}
McGregor P.,  et~al., 2004, in Moorwood A.,  Iye M.,  eds,  \procspie Vol.
  5492, Ground-based Instrumentation for Astronomy. pp 1033--1044,
  \mn@doi{10.1117/12.550288}

\bibitem[\protect\citeauthoryear{{Meyer}, {K{\"u}rster}, {Arcidiacono},
  {Ragazzoni}  \& {Rix}}{{Meyer} et~al.}{2011}]{Meyer2011}
{Meyer} E.,  {K{\"u}rster} M.,  {Arcidiacono} C.,  {Ragazzoni} R.,   {Rix}
  H.-W.,  2011, \mn@doi [\aap] {10.1051/0004-6361/201016053}, \href
  {http://adsabs.harvard.edu/abs/2011A%26A...532A..16M} {532, A16}

\bibitem[\protect\citeauthoryear{Mugnier}{Mugnier}{2008}]{Mugnierbook}
Mugnier L.,  2008, in L\'ena P.,  Rouan D.,  Lebrun F.,  Mignard F.,   Pelat
  D.,  eds, , L'observation en astrophysique.
EDP Sciences, Les Ulis, France, Chapt. 9, section~6, pp 591--613

\bibitem[\protect\citeauthoryear{Mugnier, Fusco  \& Conan}{Mugnier
  et~al.}{2004}]{Mugnier2004}
Mugnier L.~M.,  Fusco T.,   Conan J.-M.,  2004, \mn@doi [Journal of the Optical
  Society of America. A, Optics, image science, and vision]
  {10.1364/JOSAA.21.001841}, 21, 1841

\bibitem[\protect\citeauthoryear{Munro}{Munro}{1995}]{Munro1995}
Munro D.~H.,  1995, \mn@doi [Comput. Phys.] {10.1063/1.4823451}, 9, 609

\bibitem[\protect\citeauthoryear{Neichel et~al.,}{Neichel
  et~al.}{2014a}]{Neichel2013a}
Neichel B.,  et~al., 2014a, Proc. AO4ELT3, pp~1--9

\bibitem[\protect\citeauthoryear{{Neichel} et~al.,}{{Neichel}
  et~al.}{2014b}]{Neichel2014a}
{Neichel} B.,  et~al., 2014b, \mn@doi [\mnras] {10.1093/mnras/stu403}, \href
  {http://adsabs.harvard.edu/abs/2014MNRAS.440.1002N} {440, 1002}

\bibitem[\protect\citeauthoryear{{Neichel}, {Lu}, {Rigaut}, {Ammons},
  {Carrasco}  \& {Lassalle}}{{Neichel} et~al.}{2014c}]{Neichel2014b}
{Neichel} B.,  {Lu} J.~R.,  {Rigaut} F.,  {Ammons} S.~M.,  {Carrasco} E.~R.,
  {Lassalle} E.,  2014c, \mn@doi [\mnras] {10.1093/mnras/stu1766}, \href
  {http://cdsads.u-strasbg.fr/abs/2014MNRAS.445..500N} {445, 500}

\bibitem[\protect\citeauthoryear{{Reid} \& {Menten}}{{Reid} \&
  {Menten}}{2007}]{Reid2007}
{Reid} M.~J.,  {Menten} K.~M.,  2007, \mn@doi [\apj] {10.1086/523085}, \href
  {http://adsabs.harvard.edu/abs/2007ApJ...671.2068R} {671, 2068}

\bibitem[\protect\citeauthoryear{{Rigaut} et~al.,}{{Rigaut}
  et~al.}{2014}]{Rigaut2014}
{Rigaut} F.,  et~al., 2014, \mn@doi [\mnras] {10.1093/mnras/stt2054}, \href
  {http://adsabs.harvard.edu/abs/2014MNRAS.437.2361R} {437, 2361}

\bibitem[\protect\citeauthoryear{{Service}, {Lu}, {Campbell}, {Sitarski},
  {Ghez}  \& {Anderson}}{{Service} et~al.}{2016}]{Service2016}
{Service} M.,  {Lu} J.~R.,  {Campbell} R.,  {Sitarski} B.~N.,  {Ghez} A.~M.,
  {Anderson} J.,  2016, \mn@doi [\pasp] {10.1088/1538-3873/128/967/095004},
  \href {http://adsabs.harvard.edu/abs/2016PASP..128i5004S} {128, 095004}

\bibitem[\protect\citeauthoryear{{Testor}, {Lemaire}, {Kristensen}, {Field}  \&
  {Diana}}{{Testor} et~al.}{2007}]{Testor2007a}
{Testor} G.,  {Lemaire} J.~L.,  {Kristensen} L.~E.,  {Field} D.,   {Diana} S.,
  2007, \mn@doi [\aap] {10.1051/0004-6361:20066926}, \href
  {http://adsabs.harvard.edu/abs/2007A%26A...469..459T} {469, 459}

\bibitem[\protect\citeauthoryear{{Trippe}, {Davies}, {Eisenhauer}, {Schreiber},
  {Fritz}  \& {Genzel}}{{Trippe} et~al.}{2010}]{Trippe2010}
{Trippe} S.,  {Davies} R.,  {Eisenhauer} F.,  {Schreiber} N.~M.~F.,  {Fritz}
  T.~K.,   {Genzel} R.,  2010, \mn@doi [\mnras]
  {10.1111/j.1365-2966.2009.15940.x}, \href
  {http://adsabs.harvard.edu/abs/2010MNRAS.402.1126T} {402, 1126}

\bibitem[\protect\citeauthoryear{{Turri}, {McConnachie}, {Stetson},
  {Fiorentino}, {Andersen}, {V{\'e}ran}  \& {Bono}}{{Turri}
  et~al.}{2015}]{Turri2015}
{Turri} P.,  {McConnachie} A.~W.,  {Stetson} P.~B.,  {Fiorentino} G.,
  {Andersen} D.~R.,  {V{\'e}ran} J.-P.,   {Bono} G.,  2015, \mn@doi [\apjl]
  {10.1088/2041-8205/811/2/L15}, \href
  {http://adsabs.harvard.edu/abs/2015ApJ...811L..15T} {811, L15}

\bibitem[\protect\citeauthoryear{{Ubeda} \& {Kozhurina-Platais}}{{Ubeda} \&
  {Kozhurina-Platais}}{2013}]{Ubeda2013}
{Ubeda} L.,  {Kozhurina-Platais} V.,  2013, Technical report, {ACS/WFC
  Geometric Distortion: a time dependency study}

\bibitem[\protect\citeauthoryear{Ye, Gao, Wang, Cheng, Wang  \& Sun}{Ye
  et~al.}{2014}]{Ye2014}
Ye J.,  Gao Z.,  Wang S.,  Cheng J.,  Wang W.,   Sun W.,  2014, \mn@doi [J.
  Opt. Soc. Am. A] {10.1364/JOSAA.31.002304}, 31, 2304

\bibitem[\protect\citeauthoryear{{Yelda}, {Lu}, {Ghez}, {Clarkson}, {Anderson},
  {Do}  \& {Matthews}}{{Yelda} et~al.}{2010}]{Yelda2010}
{Yelda} S.,  {Lu} J.~R.,  {Ghez} A.~M.,  {Clarkson} W.,  {Anderson} J.,  {Do}
  T.,   {Matthews} K.,  2010, \mn@doi [\apj] {10.1088/0004-637X/725/1/331},
  \href {http://adsabs.harvard.edu/abs/2010ApJ...725..331Y} {725, 331}

\bibitem[\protect\citeauthoryear{{d'Orgeville} et~al.,}{{d'Orgeville}
  et~al.}{2012}]{DOrgeville2012}
{d'Orgeville} C.,  et~al., 2012, in Adaptive Optics Systems III. p. 84471Q,
  \mn@doi{10.1117/12.925813}

\makeatother
\end{thebibliography}
\bsp	% typesetting comment
\label{lastpage}
\end{document}